\documentclass{ws-mpla}
\usepackage[super]{cite}

\usepackage{amsmath,amssymb}
\usepackage{epsfig}
\usepackage{color}
\newcommand{\n}{\nonumber}
\newcommand{\be}{\begin{equation}}
\newcommand{\ee}{\end{equation}}
\newcommand{\bea}{\begin{eqnarray}}
\newcommand{\eea}{\end{eqnarray}}

\begin{document}

%\rightline{22 Apr 2022}

\title{Quantum chaos and  H\'enon-Heiles model: \\Dirac's variational approach with Jackiw-Kerman function}

\author{Choon-Lin Ho}
%\affiliation
\address{Department of Physics, Tamkang University, Tamsui 251, Taiwan}
 
\author{Chung-I Chou}
%\affiliation
\address{Department of Physics, Chinese Culture University, Taipei 111, Taiwan}

\maketitle  % for WS

\begin{abstract}

A simple semiclassical H\'enon-Heiles model is constructed based on  Dirac's time-dependent variational principle.
We obtain an effective semiclassical Hamiltonian using a Hartree-type two-body trial wavefunction in the Jackiw-Kerman form. Numerical results show that quantum effects can in fact induce chaos in the non-chaotic regions of the classical H\'enon-Heiles model.

\end{abstract}

%\pacs{05.45.Mt, 05.45.-a, 03.65.Sq}
% 05.45.Mt Quantum chaos; semiclassical methods
% 05.45.-a  Nonlinear dynamics and chaos
% 03.65.Sq  Semiclassical theories and applications

%\keywords{H\'enon-Heiles model, Dirac's variational principle, Quantum Chaos, Jackiw-Kerman function}

%\maketitle  % for LaYex

%%%%%

\section{Introduction}

In recent years, interests in classical and quantum chaos have been revived owing to the interesting part they may play in issues such as quantum entanglement, quantum coherence, quantum localization,  fast information scrambling, thermalization, etc.,  in quantum many-body systems (see e.g., Ref.[1-4]), and also in quantum field theory and gravity  (see e.g., Ref.[5-9]).

Classical chaos is concerned with the  sensitivity of the dynamics of a system to its initial conditions. Two trajectories in phase space whose initial conditions are very close will diverge exponentially in time. The rate of this separation is characterized by the Lyapunov exponent.

In quantum mechanics, the notion of phase-space trajectories loses its meaning owing to the Heisenberg uncertainty principle. It appears  that most quantum systems do not exhibit exponential sensitivity and chaos. Still,  there are examples of quantum systems that show chaotic behaviors. For instance, the hydrogen atom in a strong magnetic field displays strong irregularities in its spectrum \cite{FW}, and the wave functions of the quantum mechanical model of the stadium billiard  shows irregular patterns \cite{MK}.  Hence,  it is natural to look for quantum manifestations of chaos in classically chaotic systems  \cite{Ha}. 

 Most research efforts in quantum chaos  concerns the quantization of classically chaotic systems in the semiclassical regime. 
Different approaches have been adopted to identify the signatures of chaos in these quantized systems.  
 The most commonly employed ones are: random matrices \cite{Ha, Me},  energy level dynamics \cite{Dy}, periodic orbit expansions \cite{Gu}, Gaussian wave-packet dynamics \cite{He1,He2}, etc.
 
 As semiclassical dynamics are generally believed to be qualitatively similar to those of the quantum system and the classical limit of it, so an integrable classical system would not be chaotic in its semiclassical approximation.  That is, quantum fluctuations will suppress chaos. However, in [18] it has been demonstrated that  this is not always true. 
Using a semiclassical dynamics derived via the Ehrenfest theorem, 
the authors showed that  for the double-well system, quantum fluctuations may induce chaos.  More recently, while we were preparing this manuscript,  a new work \cite{Ber} came to our attention which shows that 
quantum corrections by metric extensions also favor chaotic behavior in the dynamics of a probe particle near the horizon of a generalized Schwarzschild black hole. 

 We would like to examine if there are other systems in which chaos could be induced by quantum effects. The example we consider is the H\'enon-Heiles model \cite{HH}.   Originally the H\'enon-Heiles potentiall was used to model the motion of a star in  the gravitation field of a galaxy, but later it was found to be also useful as a model of triatomic molecule in quantum chemistry. Thus it becomes of interest to study the quantum behaviors of the H\'enon-Heiles model.  

It is known that the H\'enon-Heiles potential admits both regular and chaotic motions ([see, e.g., Ref. [21-23]).
In Ref.[24], by determining the quantum energy levels of the H\'enon-Heiles system, it was found that  the energy levels  in the classically quasi-periodic regime continued smoothly into the classically stochastic regime.  Thus quantum fluctuations appear to suppress classical chaos in the H\'enon-Heiles potential.   More recently, studies of a semiclassical H\'enon-Heiles model using the method of Gaussian effective potential  also indicate that quantum fluctuations destroy the chaotic behavior in  the H\'enon-Heiles potential \cite{CS,PS2}.
 
In this work, we propose to study a semiclassical H\'enon-Heiles model based  on Dirac's time-dependent  variational principle \cite{Di,Fr}.   In this approach, one first constructs the effective action
$\Gamma=\int dt
~\langle \Psi,t|\,i\hbar\,\partial_t -{\cal H}|\Psi,t\rangle$ for a given system
described
by a Hamiltonian $\cal H$ and a quantum state $|\Psi,t\rangle$ parametrized by some time-dependent $c$-variables.  Variation of $\Gamma$ is then the  quantum
analogue of the Hamilton's principle. This gives the Hamiltonian equations for the $c$-variables.   The time-dependent Hartree 
approximation emerges when a specific ansatz is made for the state
$|\Psi,t\rangle$.

For our semiclassical H\'enon-Heiles model, we shall assume 
 the two-body state  $|\Psi,t\rangle$ to be factorisable into single-particle states described by
the Jackiw-Kerman (JK) function \cite{JK,TF,Ho}.  
With the time-dependent $c$-variables in these JK functions, the number of degrees of freedom of the semiclasscial system is twice that of the classical one. The new variables could introduce additional nonlinearity into the system.  And this may induce chaos in the originally regular regime of the classical system.  We investigate such possibility by numerically varying the effective Planck constant (to be defined in Sect.\,2) in the effective action.  We find that this is indeed the case.  Thus, our semiclassical H\'enon-Heiles model shows that quantum effects could induce chaos in classically non-chaotic systems.

After this work was submitted, we became aware of Ref.\,[32], where the same approach was applied to the study of semiquantum chaos in the one-dimensional double-well oscillator model.  The study of this simple model, as explained by the authors, was to serve as a first step towards a better understanding of the $(3+1)$-dimensional  scalar Higgs field model \cite{E1,E2}. Unlike the H\'enon-Heiles model, which admits both regular and chaotic motions, the classical double-well oscillator system behaves regularly.  Using as control parameter the total energy of the semiclassical system, determined by the $c$-variables in the initial JK state function,  it was also realized that  there are energy dependent transitions between regular and semiquantum chaos. 

We shall derive in Sect.\,2 the effective Hamiltonian of the semiclassical H\'enon-Heiles model based on the Dirac's variational principle.  Dynamical behaviors of the corresponding classical model are  discussed in Sect.\,3, and those of the semiclassical system in  Sect.\,4.  Sect.\,5 concludes the paper.

%------------------------------------

\section{The H\'enon-Heiles model}

The Hamiltonian of the quantum H\'enon-Heiles model is
\be
{\cal H}=\frac12 \sum_{i=1}^2\left({\hat p}_i^2+ {\hat x}_i^2\right)+ \lambda\left(\hat{x}_1^2 \hat{x}_2-\frac13 \hat{x}_2^3\right),
\label{CHH}
\ee
where $\hat{x}_i$ and $\hat{p}_i=-i\hbar\partial/\partial x_i$ are the position and momentum operators of the $i$th particle, and $\lambda$ is the coupling strength.
To study quantum effects one might vary the Planck constant $\hbar$, but this may leave an uneasy feeling that $\hbar$ is a natural constant and thus  cannot be changed.  As such, to facilitate numerical analysis, we find it convenient to define the model in terms of the rescaled variables
\be
\hat{Q}_i\equiv \lambda\, \hat{x}_1,~~~ \hat{P}_i\equiv \lambda\, \hat{p}_i=-i\hbar^\prime \frac{\partial}{\partial Q_i},~~ i=1,2,
\ee
where the effective Planck constant is $\hbar^\prime\equiv \lambda^2\,\hbar$.  Also we have 
 ${\cal H}=H/\lambda^2$ with
\be
H=\frac12 \sum_{i=1}^2\left({\hat P}_i^2+ {\hat Q}_i^2\right)+ {\hat Q}_1^2 {\hat Q}_2-\frac13 {\hat Q}_2^3.
\label{QHH}
\ee

As mentioned in the Introduction, 
to study quantum effect on the dynamical behavior, especially on the chaotic behavior,  of the system, 
we will  construct a semiclassical version of the H\'enon-Heiles model.
We adopt here the Dirac's time-dependent variational principle  by considering the effective action
\bea
\Gamma &=&\int dt
~\langle \Psi,t|\,i\hbar\,\partial_t -{\cal H}|\Psi,t\rangle\n\\
&=&\frac{1}{\lambda^2}\int dt~\langle \Psi,t|\,i\hbar^\prime\,\partial_t - H|\Psi,t\rangle.
\eea
We  assume the trial wavefunction of the quantum H\'enon-Heiles system to have the Hartree
form $|\Psi,t\rangle=\prod_i |\psi_i,t\rangle$, where the
normalized single-particle
state $|\psi_i,t\rangle$ is taken to be the JK wavefunction \cite{JK}:
\bea
\langle Q_i|\psi_i,t\rangle&=&\frac{1}{(2\pi\hbar^\prime
G_i)^{1/4}}   
\times \exp\Biggl\{
-\frac{1}{2\hbar^\prime}\left(Q_i-x_i\right)^2\Bigl[\frac{1}{2}G_i^{-1}
-2i \Pi_i\Bigr]+\frac{i}{\hbar^\prime}p_i\left(Q_i-x_i\right)
\Biggr\}~.\n
\\
\label{JK}
\eea
The real quantities $q_i(t)$, $p_i(t)$, $G_i(t)$ and $\Pi_i (t)$ are
variational parameters which do not varied at $t=\pm\infty$.  The JK wavefunction can be viewed as the $Q$-representation
of the squeeze state \cite{TF}.
We prefer to use the JK form since the physical meanings of the variational
parameters  in the JK wavefunction are most transparent, as we
shall show below.  Furthermore, the JK form is in the general Gaussian form so
that integrations are most easily performed.

It is not hard to work out the following expectation values:
\bea
\langle \Psi |{\hat Q}_i|\Psi\rangle =x_i, \quad\quad && 
(\Delta Q_i)^2\equiv \langle \Psi |({\hat Q}_i-x_i)^2|\Psi\rangle=\hbar^\prime G_i,\n\\
\langle \Psi |{\hat P}_i|\Psi\rangle=p_i, \quad\quad &&
(\Delta P_i)^2\equiv \langle \Psi |({\hat P}_i-p_i)^2|\Psi\rangle=4\hbar^\prime\,G_i\Pi_i^2 + \frac{\hbar^\prime}{4G_i}.
\label{EV}
\eea
It is clear that $x_i$ and $p_i$ are the expectation values of the operators
${\hat Q}_i$ and ${\hat P}_i$.  Also, $\hbar^\prime G_i$ is the mean fluctuation of the position of the $i$th particle
and that $G_i>0$.  $\Pi_i$ is related to the mean fluctuation of ${\hat P}_i$.
The uncertainty relation is 
\be
\Delta Q_i\,\Delta P_i = \frac{\hbar^\prime}{2} \sqrt{1 + (4 G_i \Pi_i)^2}.
\label{UR}
\ee 
Other expectation values needed to evaluate the effective action are:
\bea
\langle \Psi |{\hat Q}_i^2|\Psi\rangle=x_i^2  + \hbar^\prime\, G_i\ , 
\n\\
\langle \Psi |{\hat P}_i^2|\Psi\rangle= p_i^2 + 4\hbar^\prime\, G_i\Pi_i^2  + \frac{\hbar^\prime}{4G_i}\ ,
\n\\
\langle \Psi |{\hat Q}_2^3|\Psi\rangle= x_2^3  + 3\hbar^\prime\, G_2 x_2\ ,
\n\\
\langle \Psi |{\hat Q}_1^2 {\hat Q}_2|\Psi\rangle=x_1^2 x_2+ \hbar^\prime\, G_1 x_2\ ,
\n\\
\langle \Psi |i\hbar^\prime\partial_t|\Psi\rangle=\sum_i
(p_i\dot{x}_i-\hbar^\prime\,G_i\dot{\Pi}_i)\ .
\eea

With these expectation values, the effective action $\Gamma$ for
the Hamiltonian $H$ can be worked
out to be 
\be
\Gamma (x,p,G,\Pi)=\frac{1}{\lambda^2}
\int dt ~\left[\sum_i
(p_i\dot{x}_i+\hbar^\prime\Pi_i
\dot{G}_i)-H_{eff}\right],  
\label{action}
\ee
where $H_{eff}=\langle\Psi|H|\Psi\rangle$ is the effective
Hamiltonian given by
\bea
H_{eff}&=&
\frac12 \sum_{i=1}^2\left(p_i^2+ x_i^2\right)+ x_1^2x_2-\frac13 x_2^3\n\\
&&+\hbar^\prime\left[\frac12 \sum_i \left(\frac{1}{4G_i} + G_i
+4 G_i \Pi_i^2 \right) +\left(G_1-G_2\right)\,x_2\right].
\label{Heff}
\eea
 One sees from the form of the effective action $\Gamma$
that $\Pi_i$ is the canonical conjugate of $\hbar^\prime G_i$.
 The second line of (\ref{Heff}) gives the quantum contribution to the classical Hamiltonian in this semiclassical model.

%-----------------------------------
\section{The  classical system}   

From Eq.\,(\ref{Heff}), the Hamiltonian of the classical H\'enon-Heiles model is taken to be $H_c\equiv H_{eff}(\hbar^\prime=0)$, i.e.,
\be
H_c=\frac12 \sum_{i=1}^2\left(p_i^2+ x_i^2\right)+ x_1^2x_2-\frac13 x_2^3,
\label{CHH}
\ee
where $x_i$ and $p_i$ are the position and momentum of the $i$th particle.
The Hamiltonian equations of motion are
\bea
\dot{x}_1=p_1, && \quad\quad \dot{p}_1=-x_1-2 x_1x_2,\n\\
\dot{x}_2=p_2, && \quad\quad \dot{p}_2=-x_2-x_1^2 + x_2^2.
\label{cEOM}
\eea
Here  the dot represents derivative with respect to time $t$.

The behavior of this system has been well studied \cite{LL,Gu,Be,Em}.  
Fig.\,1 depicts the three-dimensional and the contour plot of the potential.   It has a three-fold rotational symmetry, 
is unbounded from below, but has a local minimum in the center within which a particle  can be confined.
It is found that the system is practically integrable for energy  below $E=1/12\approx 0.08333$, and as the energy increases the system becomes more and more ergodic, with invariant curves and ergodic regions coexisting,  and is completely ergodic at the escape energy $E=1/6\approx 0.16667$. 

Two commonly used methods to study the behavior of the system are the Poincar\'e section (or Poincar\'e surface of section) and the Lyapunov exponent. 

For Poincar\'e section, one plots the points of intersection of the orbit of the motion and  a two-dimensional plane, here taken to be  the $p_2$-$x_2$ plane with $x_1=0$ and $p_1>0$.  For simplicity, we choose $x_1(0)=x_2(0)=x_0$ and $p_1(0)=p_2(0)=p_0$.  
 The system of equations (\ref{cEOM}) is solved using a fourth-order explicit Runge-Kutta 
 method with fixed time step size of 0.02, up to total time of 20,000 units.

The Lyapunov exponent $\lambda (t)$ measures how fast two initially nearby orbits are separated as time passes.
We shall be interested in the separation of two neighboring orbits in the configuration space.                  
So we define the Lyapunov exponent  by
\be
\lambda(t)\equiv \frac{\ln d(t) -\ln d(0)}{t},
\ee
where $d(t)$ is the separation of two  nearby initial points in the configuration space at time $t$.
The system is said to be chaotic if $\lim_{t\to \infty} \lambda (t)>0$.
We take two nearby initial points $(x_1(0), x_2(0),p_1(0), p_2(0))$ and 
$(x_1^\prime(0), x_2^\prime(0), p_1^\prime(0), p_2^\prime(0)))$ with  their separation $d(t)=\sqrt{(x_1^\prime(t) -x_1(t))^2 + (x_2^\prime (t)- x_2(t))^2}$.  The two neighboring points are so chosen so that they have the same energy.  For illustration purpose, we choose  $p_1(0)=p_2(0)=p_1^\prime(0)=p_2^\prime(0)$, and $x_2^\prime (0)=x_2(0) + \Delta x_2(0)$
 with $\Delta x_2(0)=0.0001$, Then the condition of equal energy gives $x_1^\prime (0)=\sqrt{C/(2x_2^\prime(0)+1)}$, 
 where the constant $C$ is
\be
C=\left(2 x_2(0) +1\right)x_1(0)^2 + \left(x_2(0)^2- x_2^\prime(0)^2\right) -\frac23 \left(x_2(0)^3- x_2^\prime(0)^3\right) .
\label{C}
\ee

In Figs.\,2 we plot the Poincar\'e sections and Lyapunov exponents for the classical equation of motion (\ref{cEOM}) with the initial data $\{x_0, p_0\}=\{0.12, 0.001\},  \{0.10, 0.01\}$, and $ \{0.20, 0.01\}$, and in Fig.\,3 for  $p_0=0.01$ and $x_0=0.30, 0.33$, and $0.35$.
The corresponding classical energies are: $E=0.01555, 0.01077, 0.04543, 0.10810, 0.13296$ and $0.15118$, respectively. 
One notes that for energies below $E=1/12=0.08333$, the Poincar\'e sections show two invariant curves. The Lyapunov exponents are negative (signaling regular motions), or slightly positive but with regularly appearance of  negative values, indicating that the distance between the two orbits appears to have some periodic dependence.  As the energy becomes higher, such regularity fades away as ergodicity begins to set in,  ergodic regions appear in the Poincar\'e sections, and the Lyapunov exponent becomes more positive.  Similar behaviors were also  reported in [23].

%-----------------------------
\section{The semiclassical system}
 
Varying the effective action $\Gamma$ in Eq.\,(\ref{action})  with respect to $x_i,~p_i,~G_i$ and $\Pi_i$ then gives the Hamilton 
equations
of motion in the Hartree approximation:
\bea
&&\dot{x}_i=p_i,  \quad\quad  {\dot G}_i=4 G_i \Pi_i, ~~~i=1,2,
\n\\
&&\dot{p}_1=-x_1-2 x_1x_2;
\n\\
&&   \dot{p}_2=-x_2-x_1^2 + x_2^2 - \hbar^\prime \left(G_1-G_2\right),\\
\label{qEOM}
&& \dot{\Pi}_1=\frac{1}{8G_1^2} - 2\Pi_1^2 -x_2 -\frac12,
\n\\
&& \dot{\Pi}_2=\frac{1}{8G_2^2} - 2\Pi_2^2 + x_2 -\frac12.
\n
\eea
This set of equations replaces the classical equations of motion (\ref{cEOM}).

%---------------------

Our semiclassical model has an extended phase space.  To keep our model as close to the classical model as possible, we choose initial parameters so as to minimize the quantum effects.  From eq. (\ref{UR}), the uncertainty relation is minimal for $G_i(0)=0$ or $\Pi_i(0)=0$. But $G_i(0)=0$ makes the last equation in (\ref{EV}) singular,  so we take $\Pi_i(0)=0$.  To eliminate initial dependence of $x_2$ , we take $G_1=G_1$.  With these choices, the quantum part of the initial effective Hamiltonian becomes $\hbar^\prime\sum_i (G_i+1/4G_i)/2$.  This term is minimized with $G_i=1/2, i=1,2$.   Thus the initial value of the effective Hamiltonian is $H_{eff}=E +\hbar^\prime$, where $E$ is the energy evaluated using only the classical Hamiltonian $H_c$.   For $\hbar^\prime$, we take $\hbar^\prime < E/10$ in this work to keep quantum effect within reasonable bound.  Also, the choices $G_1=G_2$  and $\Pi_i=0$ mean that for the computation of the Lyapunov exponent, we can take the two neighboring initial points,  with the same effective energy, by the same criterion as in the classical case with the choice  (\ref{C}).

In Figs.\,4-7, we plot the Poincar\'e sections and Lyapunov exponents with different values of $\hbar^\prime$ for the first four sets of $\{x_0, p_0\}$ in Fig.\,2 and 3.  As discussed in Sect.\,3, these sets of parameters give regular classical motions in H\'enon-Heiles systems.   However, it is obvious that with reasonably small values of $\hbar^\prime$, the distribution  of points in the Poincar\'e section becomes more diffusive and stochastic. The corresponding  Lyapunov exponents also become more positive as $\hbar^\prime$ increases.   This indicates that, in this semiclassical approximation, quantum effects could induce chaos in classically regular systems.

It is interesting to note that, for large values of $\hbar^\prime$ considered here, the distribution of the points in the Poincar\'e sections, while appears stochastic, seems to stay in a ring-shaped region. 

%-----------------
\section{Conclusions}

In this work we have attempted to investigate, in the semiclassical approximation,  if quantum effects could induce chaos in a classically regular system.   The system we considered is the  H\'enon-Heiles model, which is known to  admit both regular and chaotic motions.  To construct a semiclassical  effective Hamiltonian that incorporates quantum correction, we have  employed the Dirac's time-dependent variational approach with a Hartree-type two-body trial wavefunction in the Jackiw-Kerman form.  The effective Hamiltonian is described by the parameters that specified the Jackiw-Kerman wavefunction.  Quantum effect on the motion of the system is studied numerically by varying the effective Planck constant in the effective Hamiltonian.   Our results show that  it is possible for quantum effects to induce chaos in classically non-chaotic systems.

It is noted that the same approach has previously  been applied to a one-body double-well oscillator model \cite{BE}.  The motion of this system is regular classically. Quantum effect on the motion of the semiclassical system was studied using the total energy as the control parameter.  It was also shown that the motion of the system can become chaotic due to  quantum effect.

%----------------------------

%%%%%%%. 
%--  
---------------------

%\acknowledgments
%\section*{Acknowledgments}
The work is supported in part by the Ministry of Science and Technology (MoST)
of the Republic of China under Grant MOST 110-2112-M-032-011.  We thank H.-Th. Elze for bringing our attention to Ref.\,[32].

%---------------------------------
\newpage

%\bibliographystyle{plain}

%---     Figures ---
%--- Fig.1
\begin{figure}[ht] \centering
\includegraphics*[width=8cm,height=8cm]{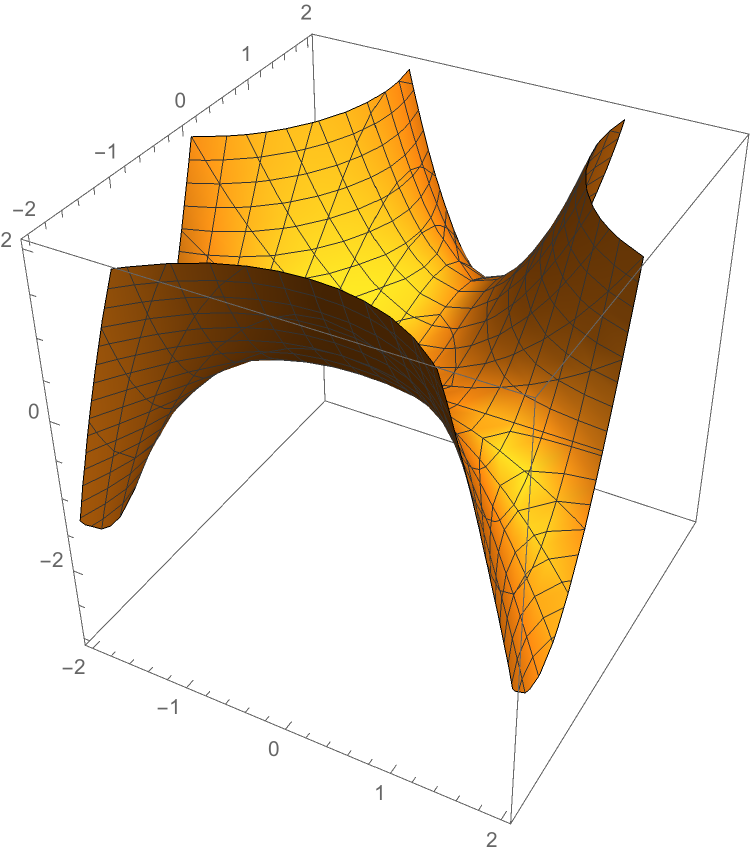}\\
\includegraphics*[width=8cm,height=8cm]{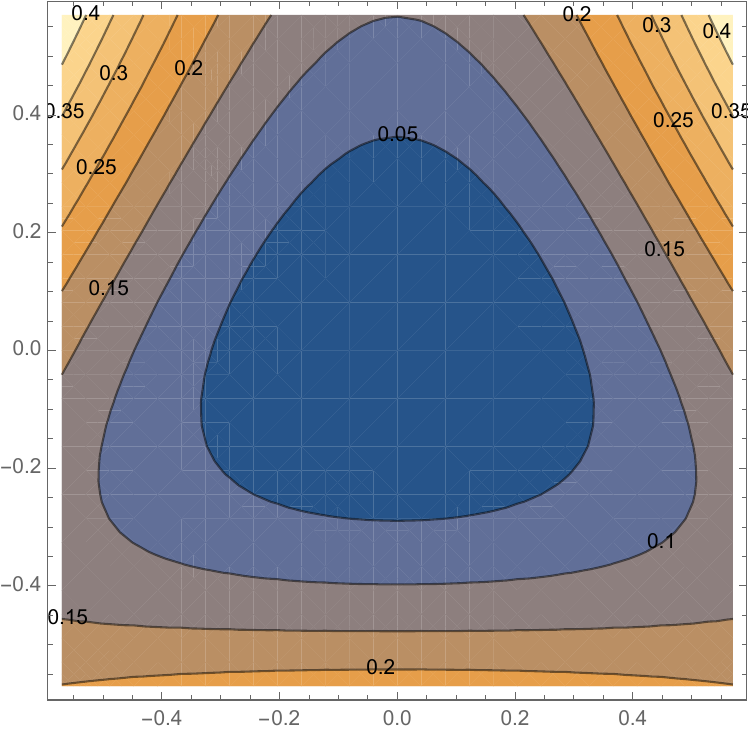}
\caption{Three-dimensional and contour plots of the classical H\'enon-Heiles potential}
\label{Fig1}
\end{figure}

% -- Fig. 2
%----------------------------------
\begin{figure}[ht] \centering
\includegraphics*[width=7cm,height=5cm]{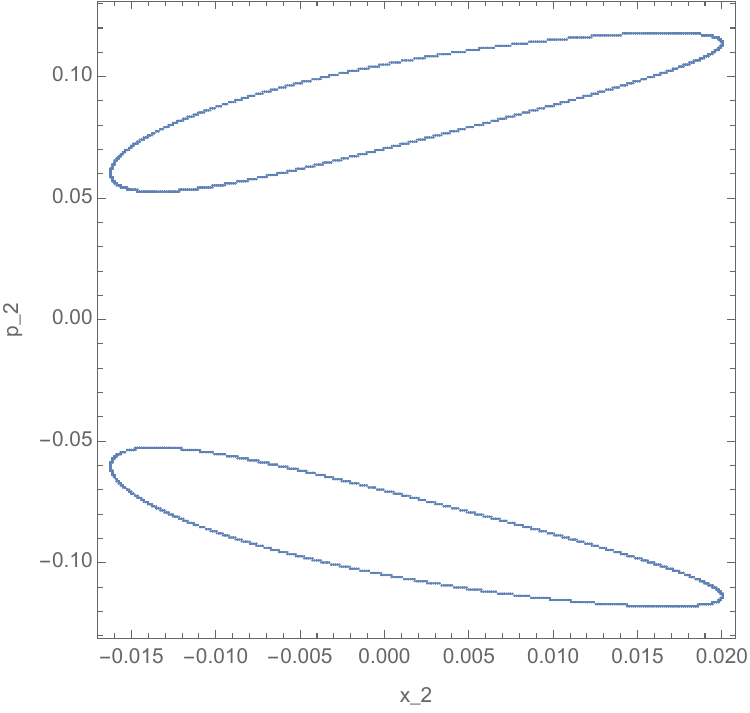}\includegraphics*[width=7cm,height=5.5cm]{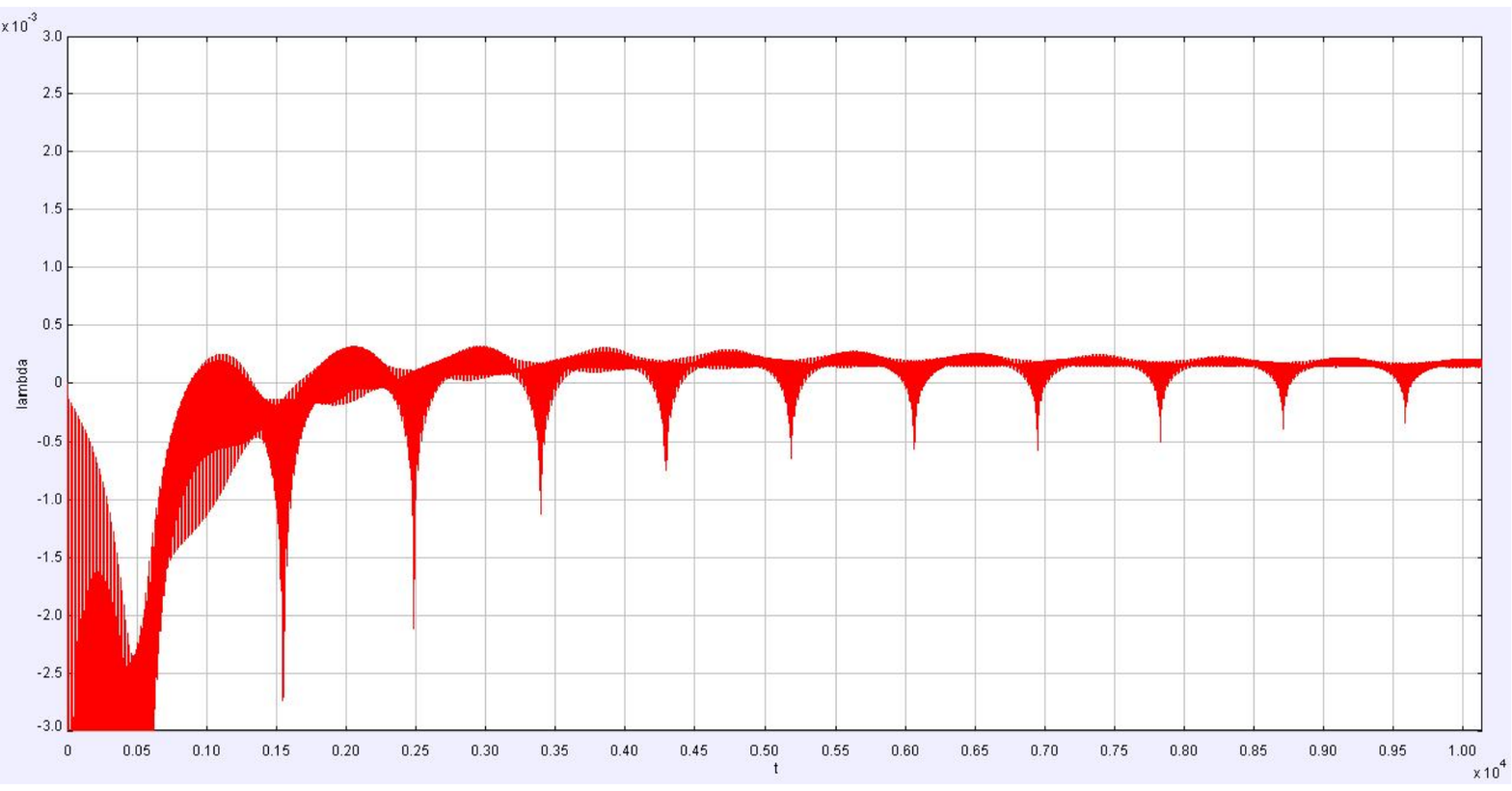}\\
\includegraphics*[width=7cm,height=5cm]{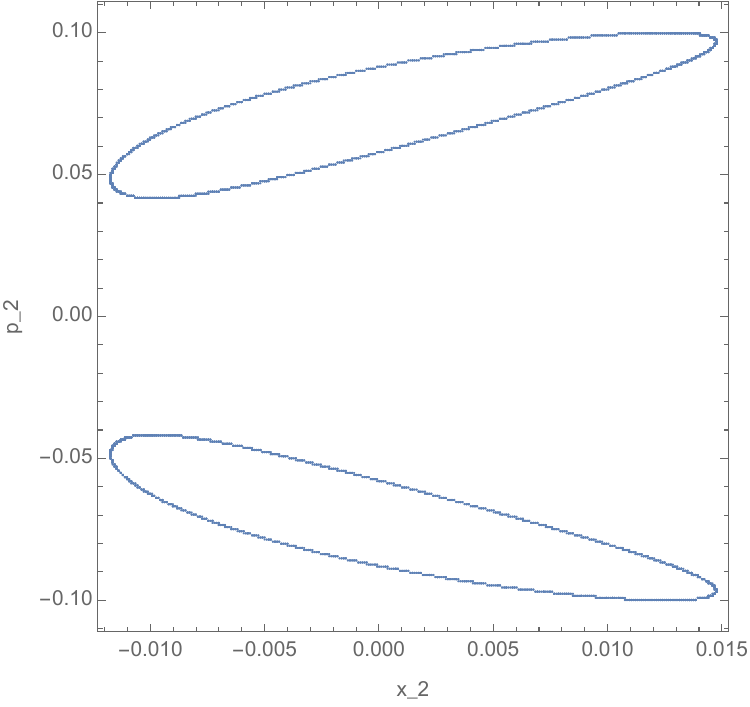}\includegraphics*[width=7cm,height=5.5cm]{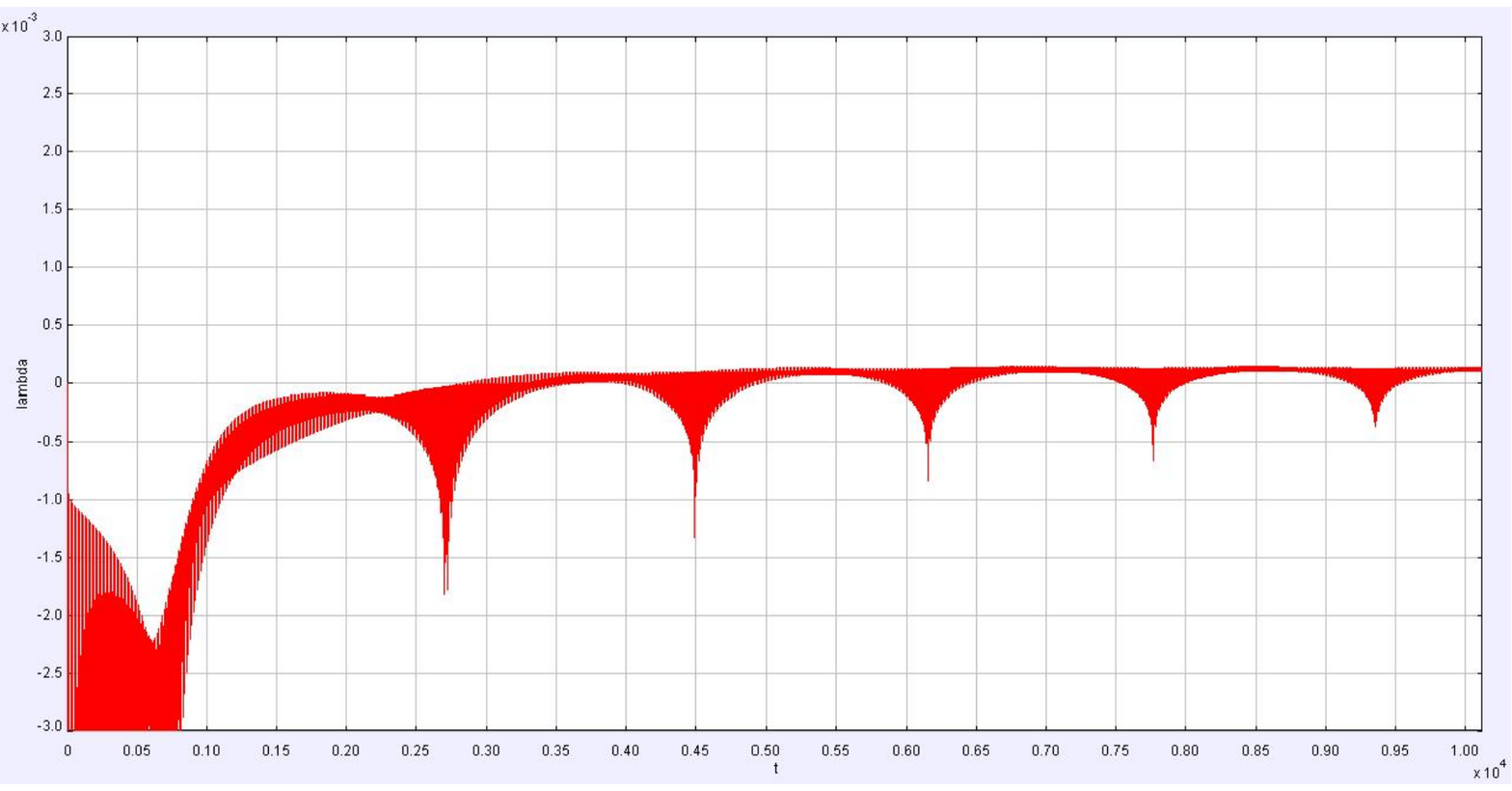}\\
\includegraphics*[width=7cm,height=5cm]{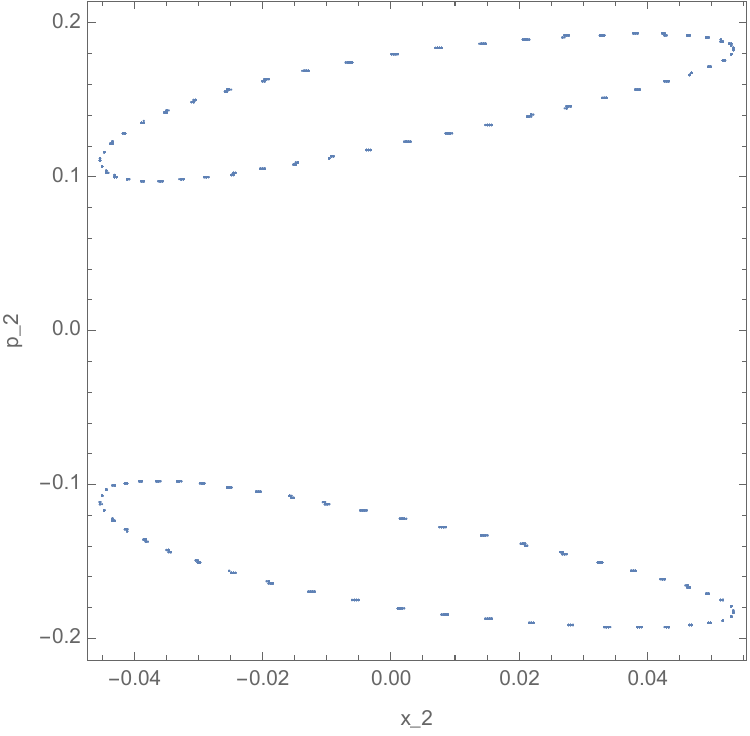}\includegraphics*[width=7cm,height=5.5cm]{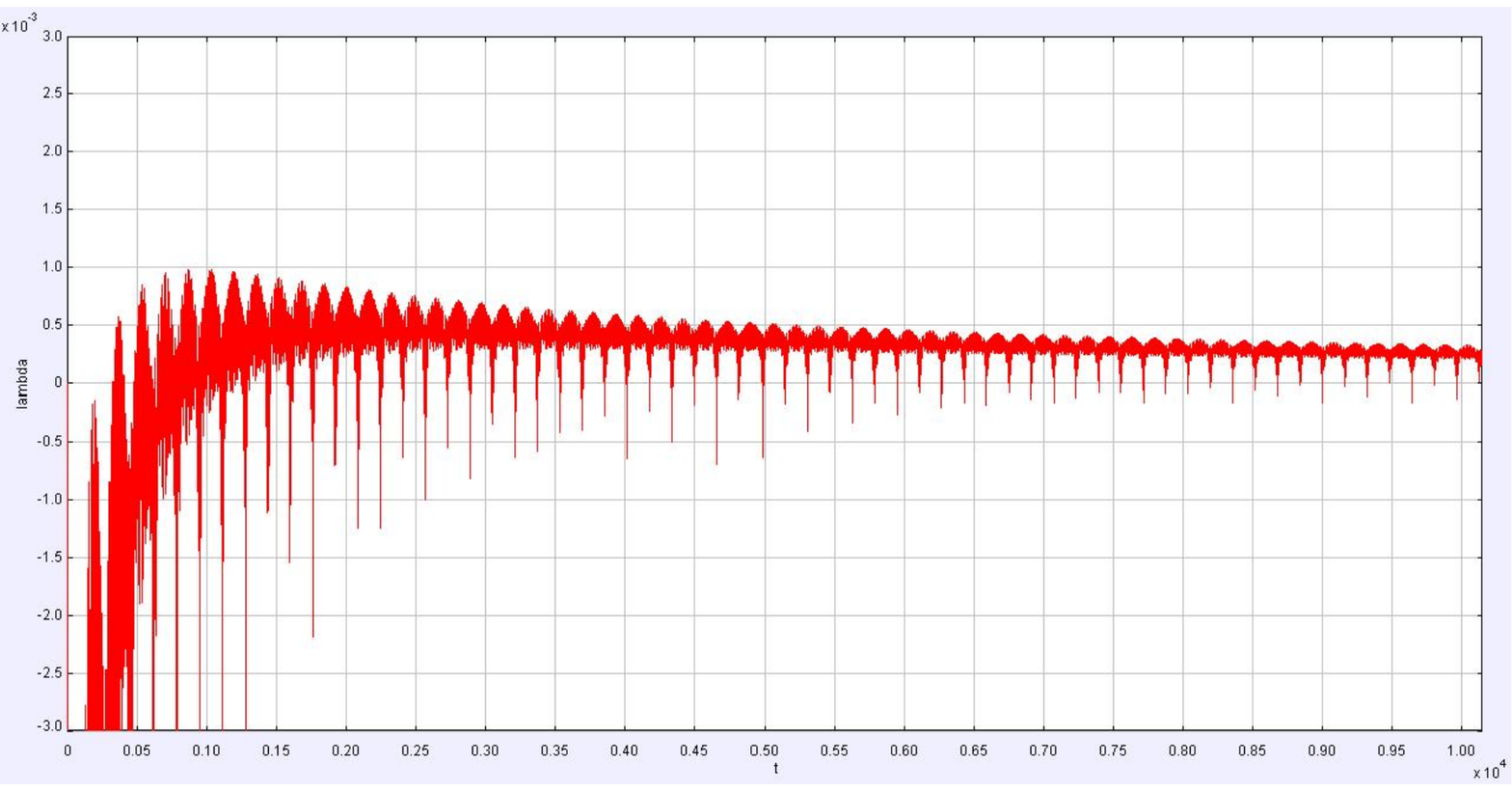}
\caption{Poincar\'e sections (left) and Lyapunov exponents (right) for the classical equation of motion (\ref{cEOM}) with the initial data $\{x_0, p_0\}=\{0.12, 0.001\},  \{0.10, 0.01\}$, and $ \{0.20, 0.01\}$ (top to bottom). The corresponding classical energies are: $E=0.01555, 0.01077$ and $0.04543$, respectively.}
\label{Fig2}
\end{figure}

% -- Fig. 3
%----------------------------------
\begin{figure}[ht] \centering
\includegraphics*[width=7cm,height=5cm]{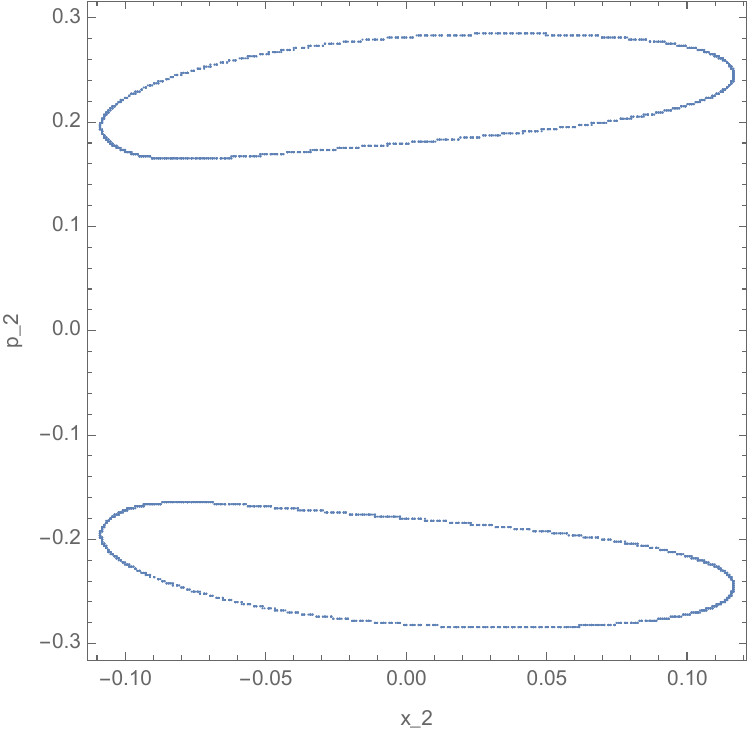}\includegraphics*[width=7cm,height=5.5cm]{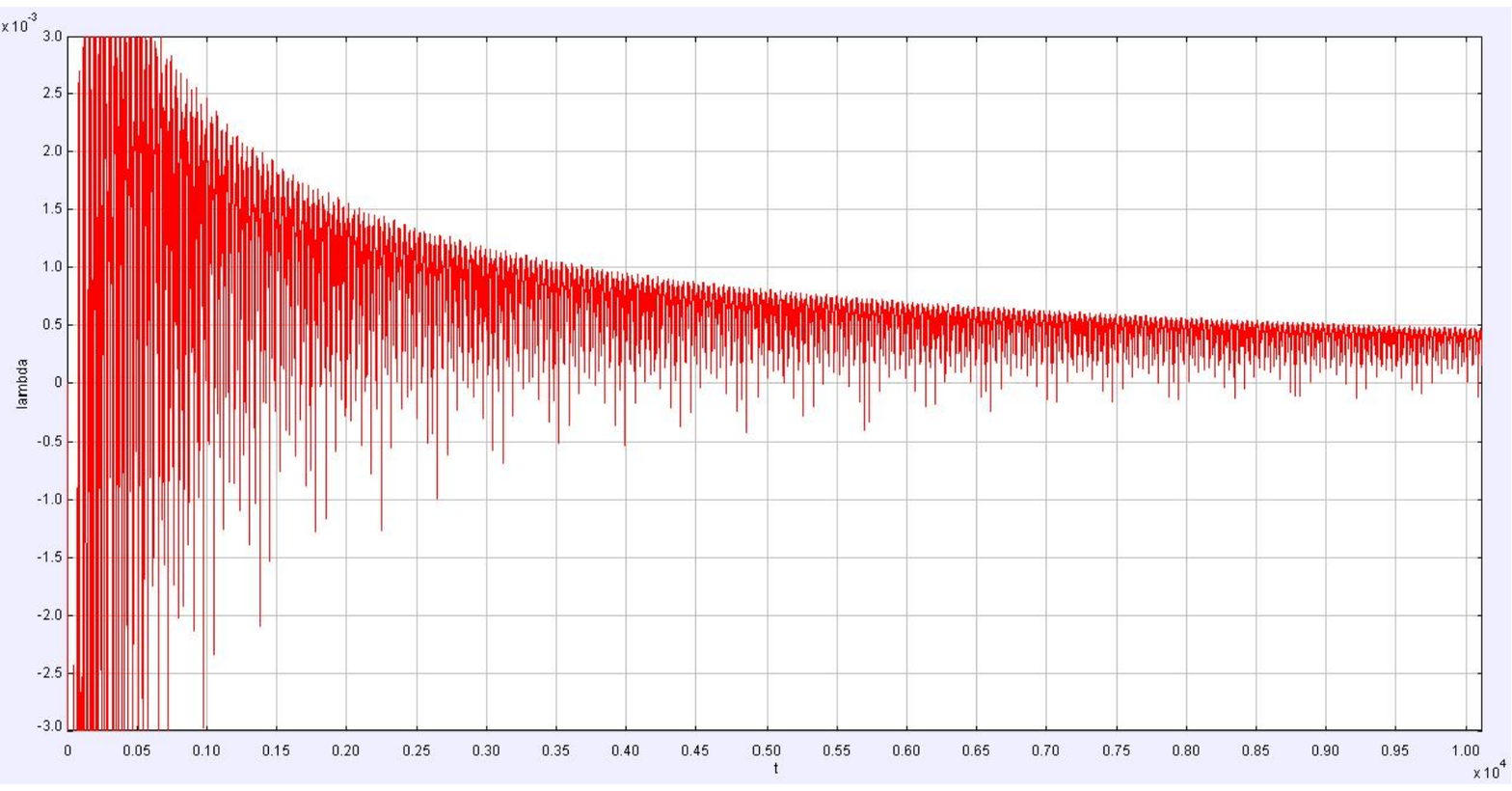}\\
\includegraphics*[width=7cm,height=5cm]{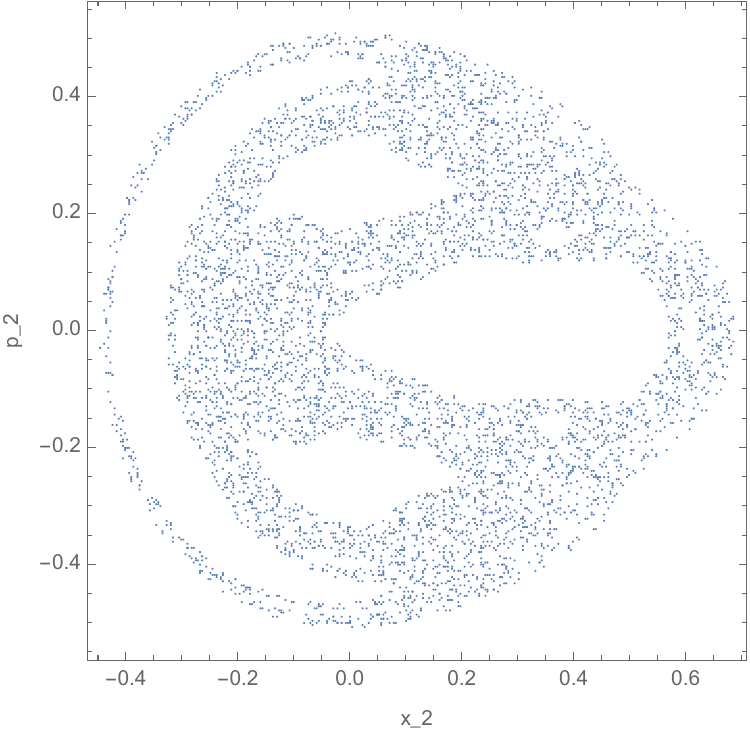}\includegraphics*[width=7cm,height=5.5cm]{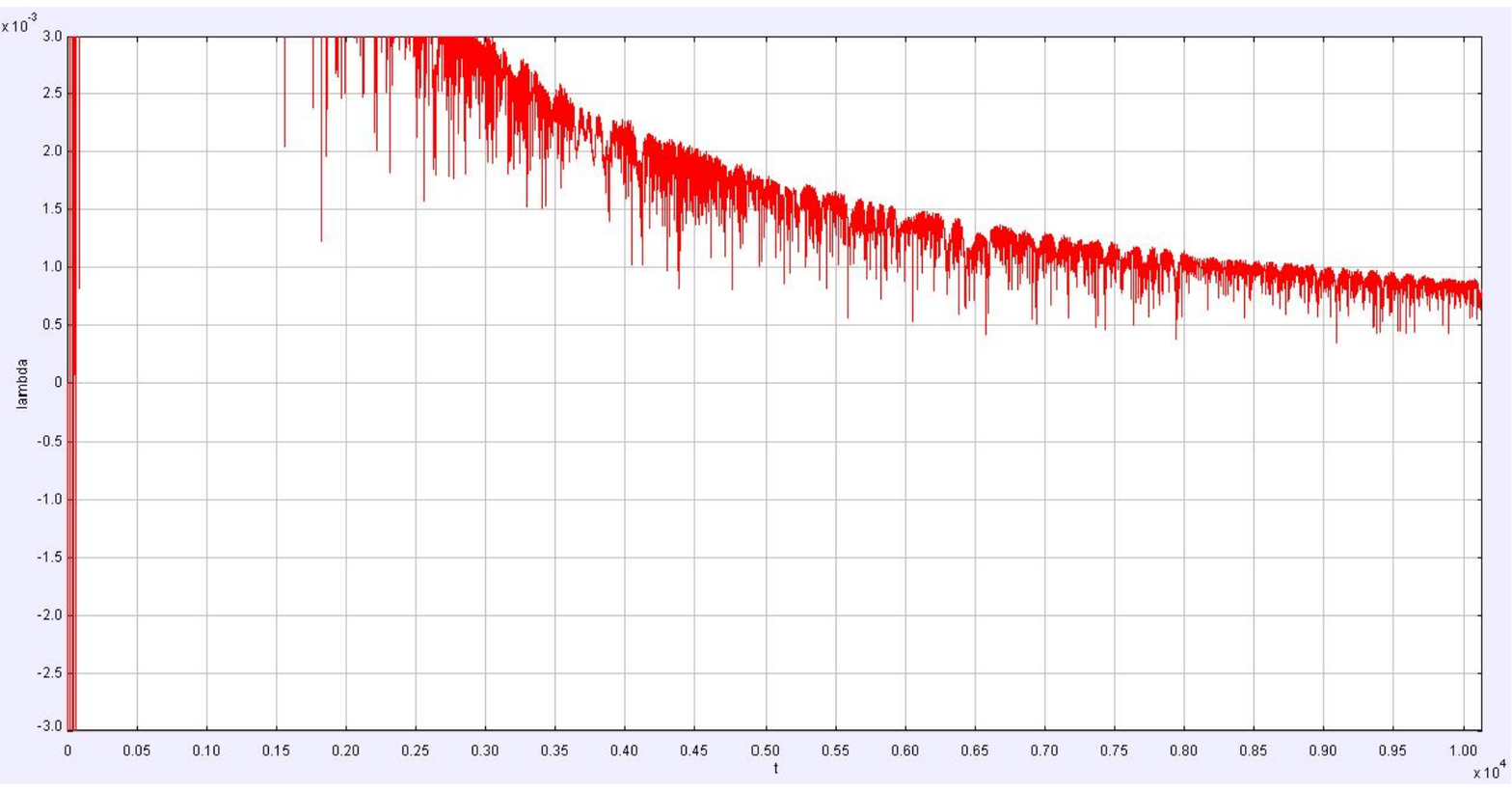}\\
\includegraphics*[width=7cm,height=5cm]{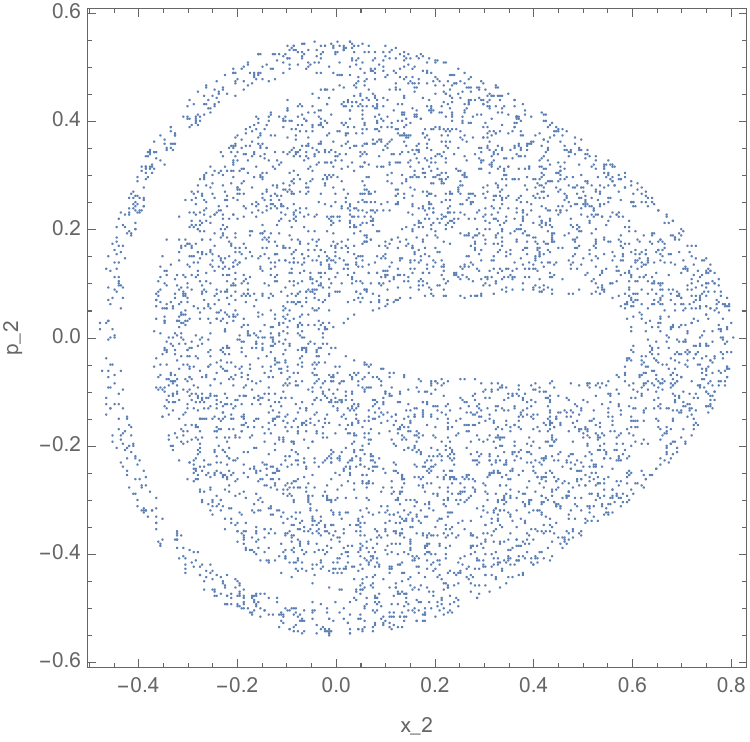}\includegraphics*[width=7cm,height=5.5cm]{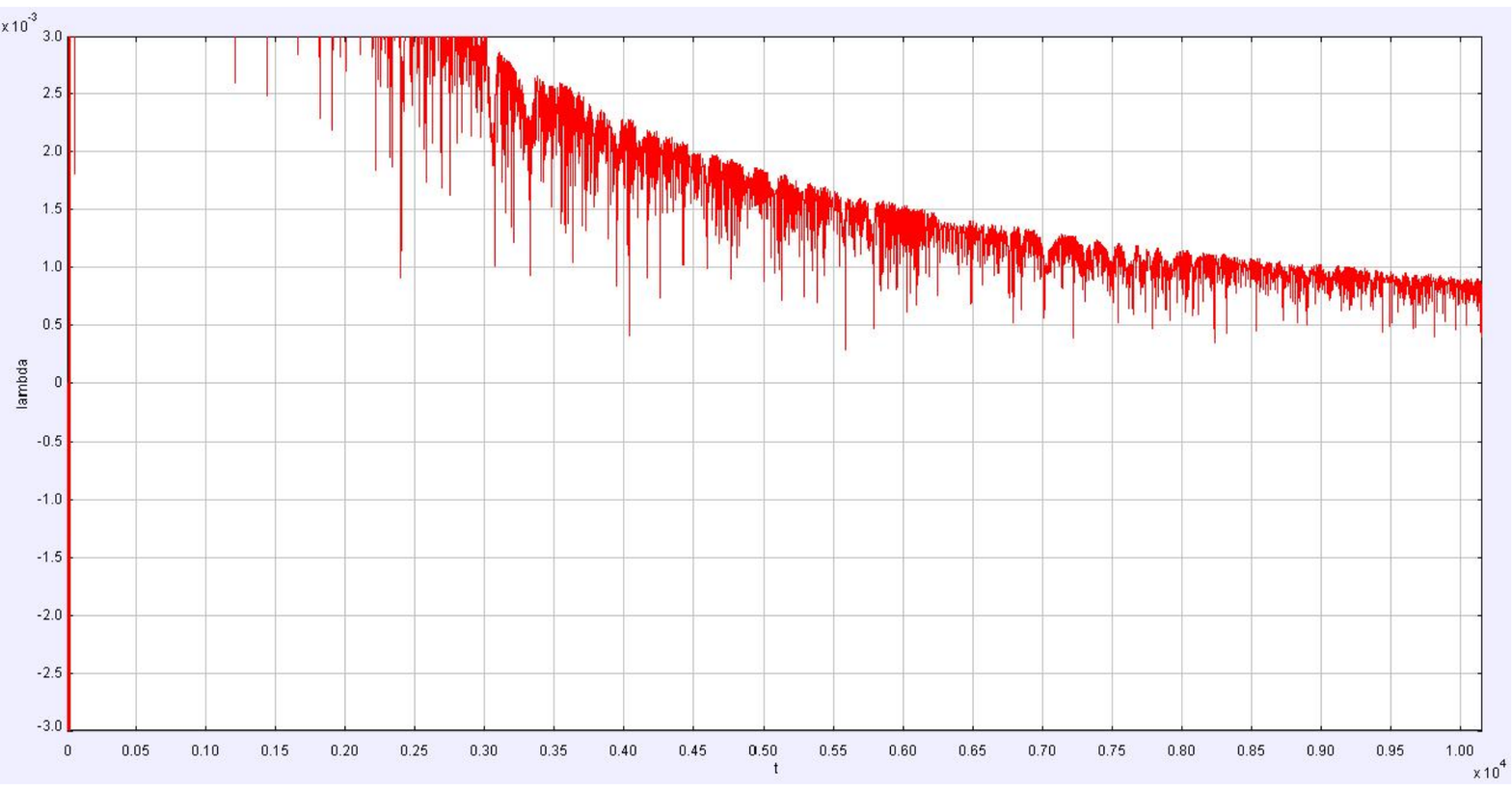}
\caption{Poincar\'e sections (left) and Lyapunov exponents (right) for the classical equation of motion (\ref{cEOM}) with the initial data $p_0=0.01$ and $x_0=0.30, 0.33$, and $0.35$ (top to bottom).  The corresponding classical energies are: $E=0.10810, 0.13296$ and $0.15118$, respectively.}
\label{Fig3}
\end{figure}

% -- Fig. 4
%----------------------------------
\begin{figure}[ht] \centering
\includegraphics*[width=7cm,height=5cm]{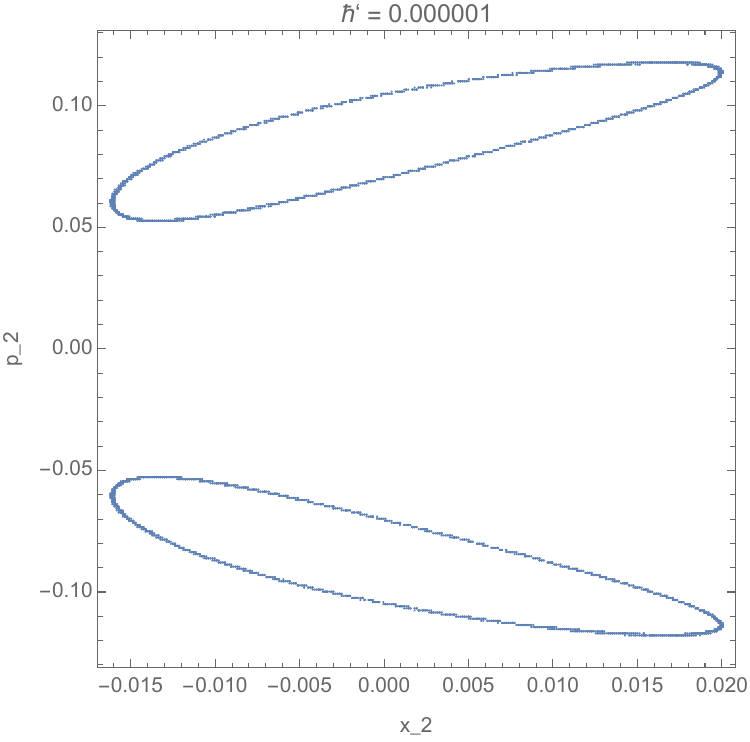}\includegraphics*[width=7cm,height=5.5cm]{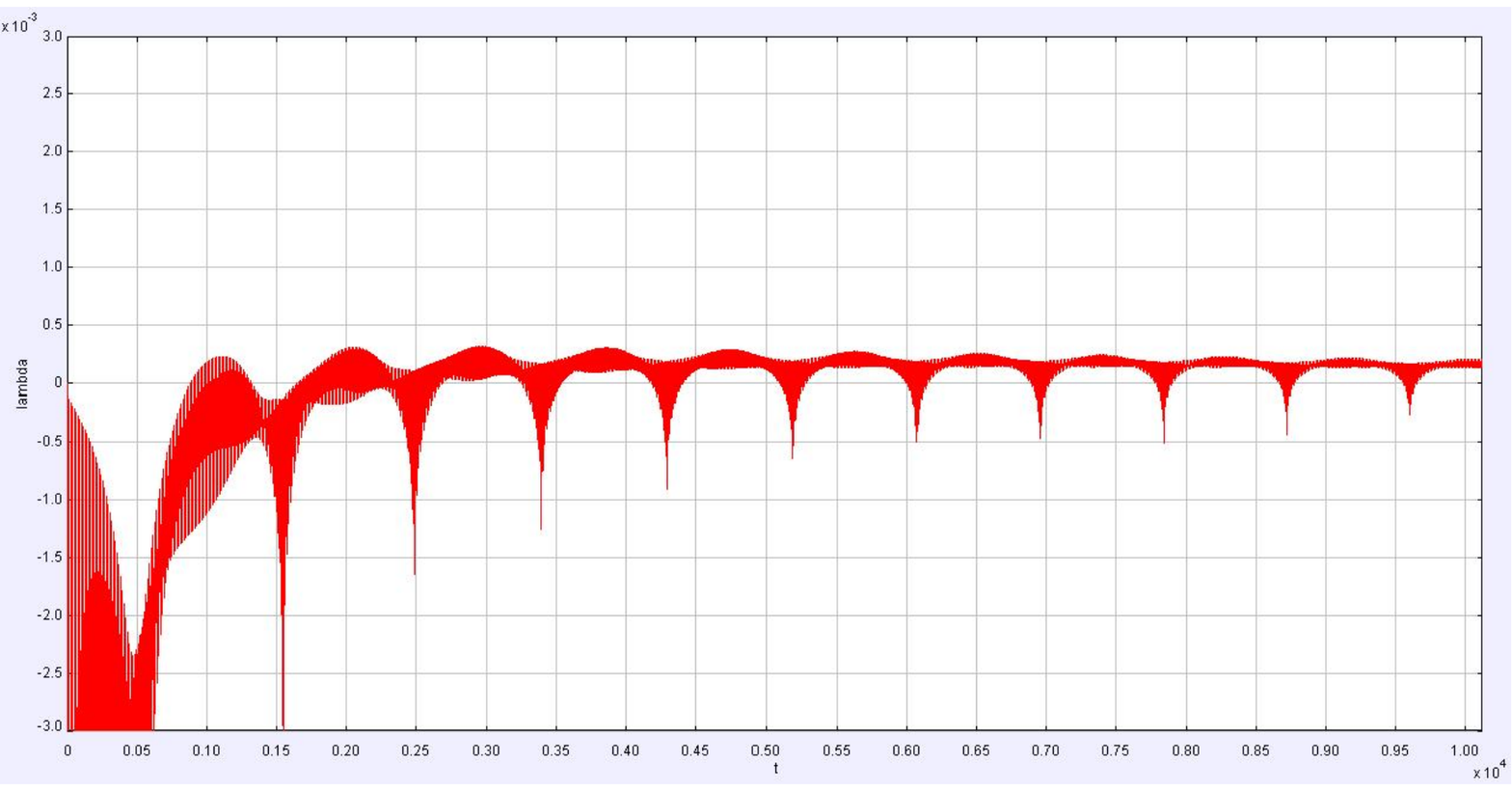}\\
\includegraphics*[width=7cm,height=5cm]{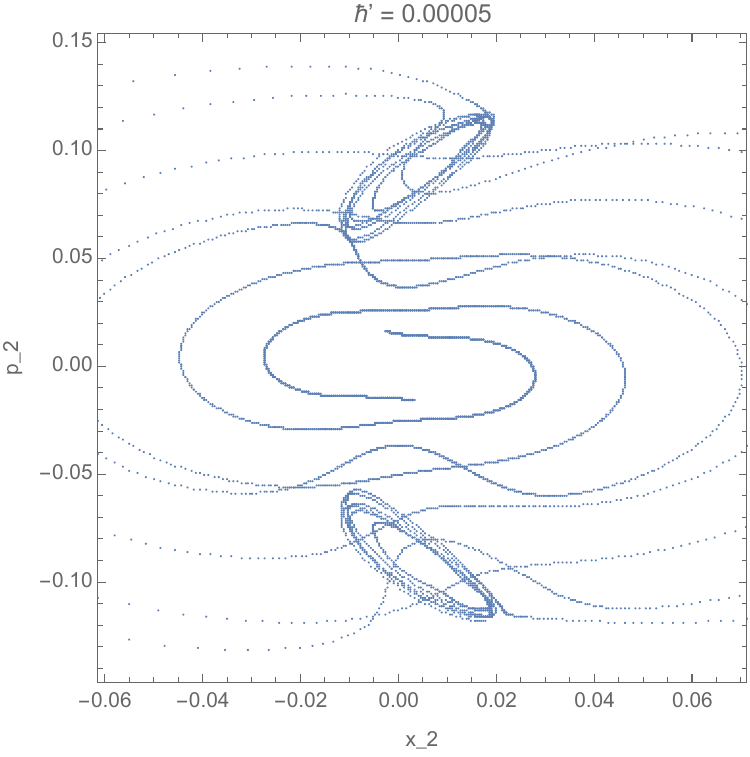}\includegraphics*[width=7cm,height=5.5cm]{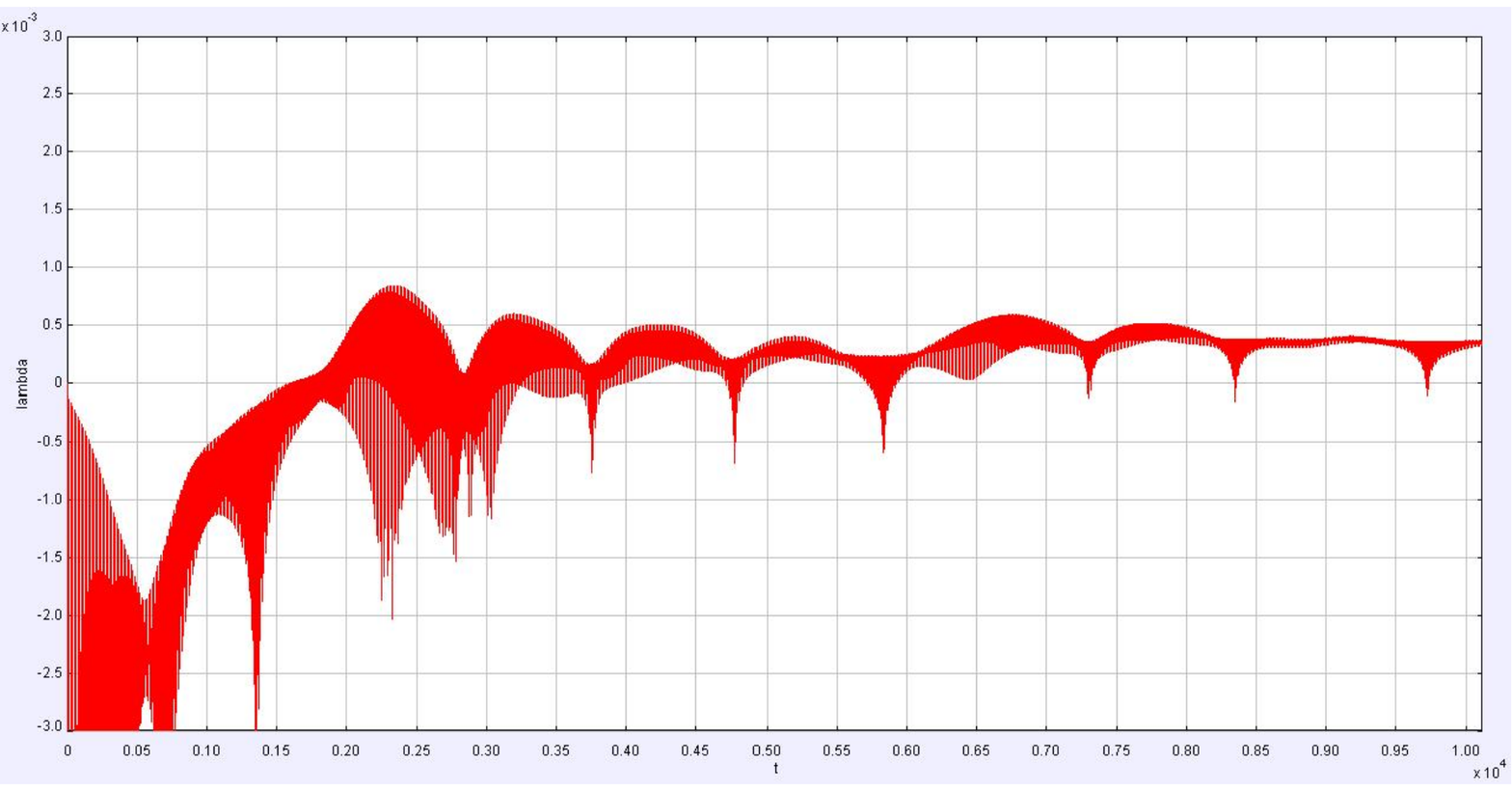}\\\includegraphics*[width=7cm,height=5cm]{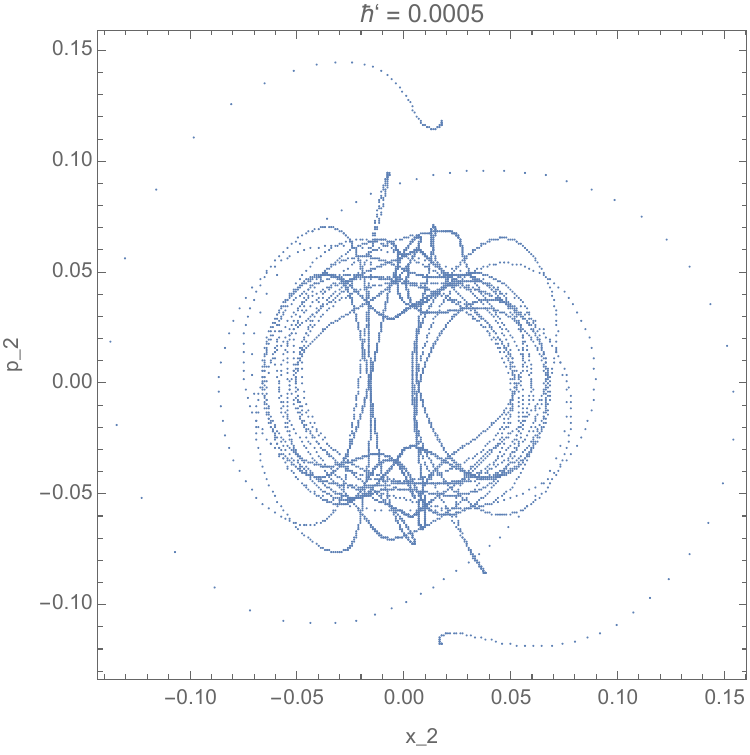}\includegraphics*[width=7cm,height=5.5cm]{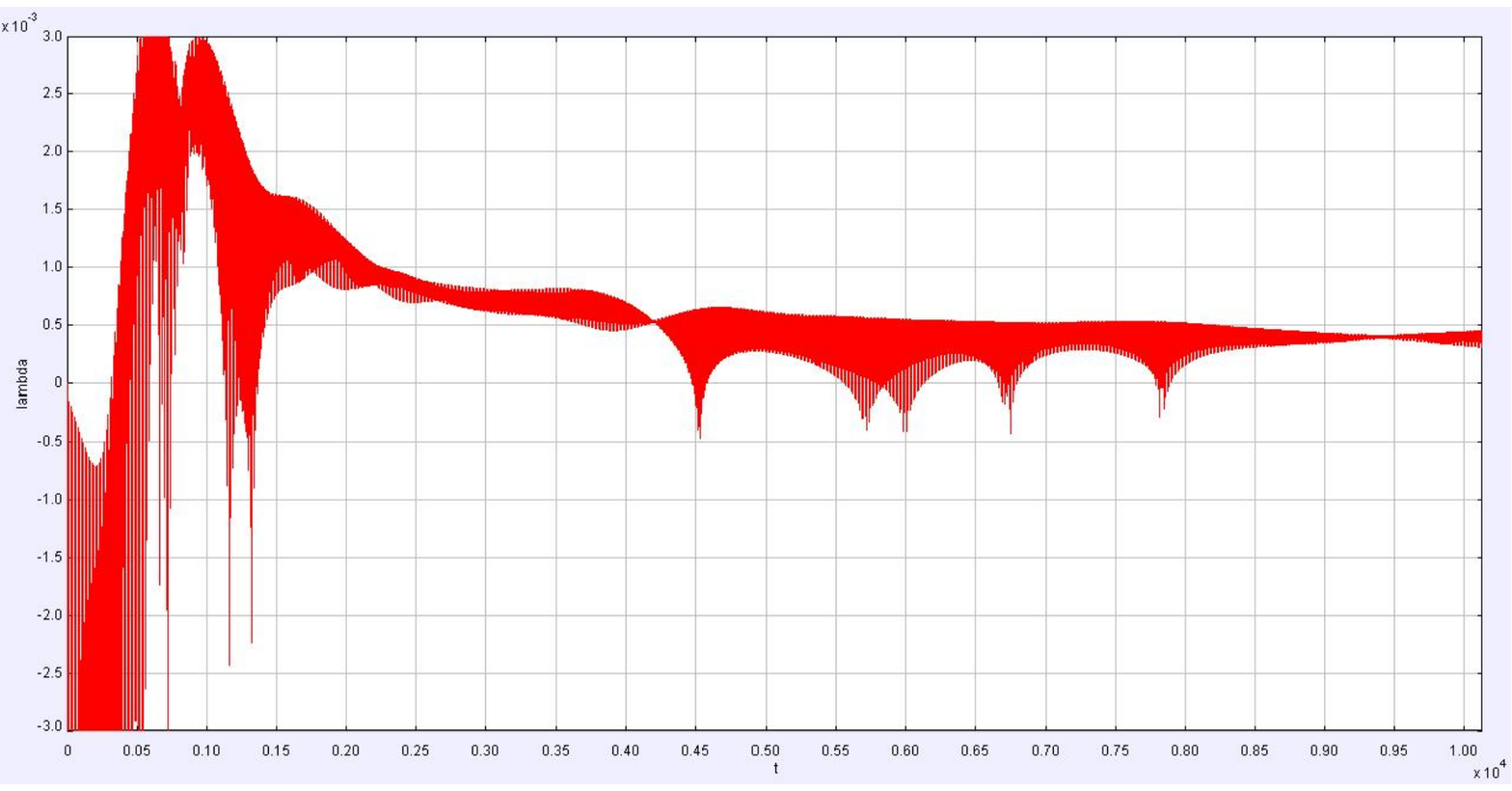}\\
\includegraphics*[width=7cm,height=5cm]{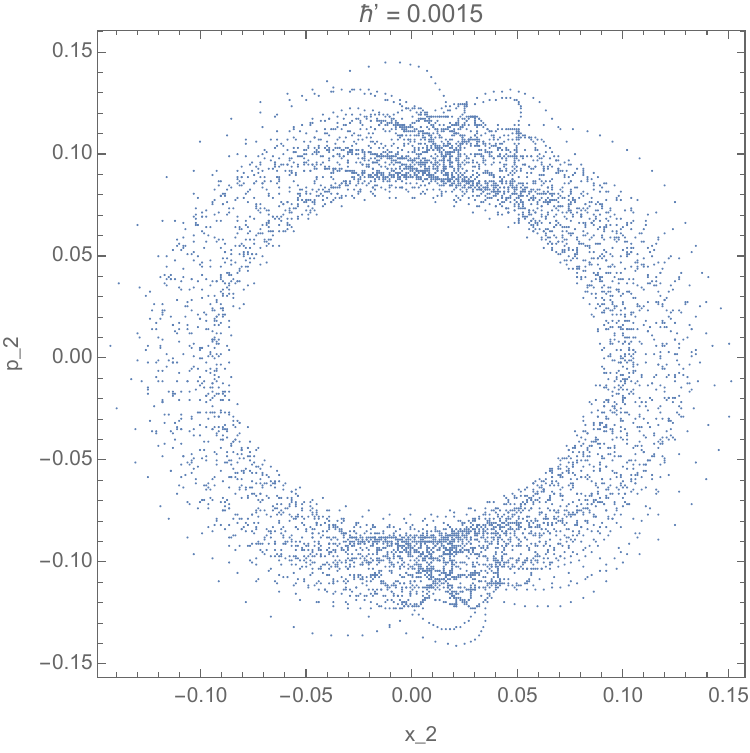}\includegraphics*[width=7cm,height=5.5cm]{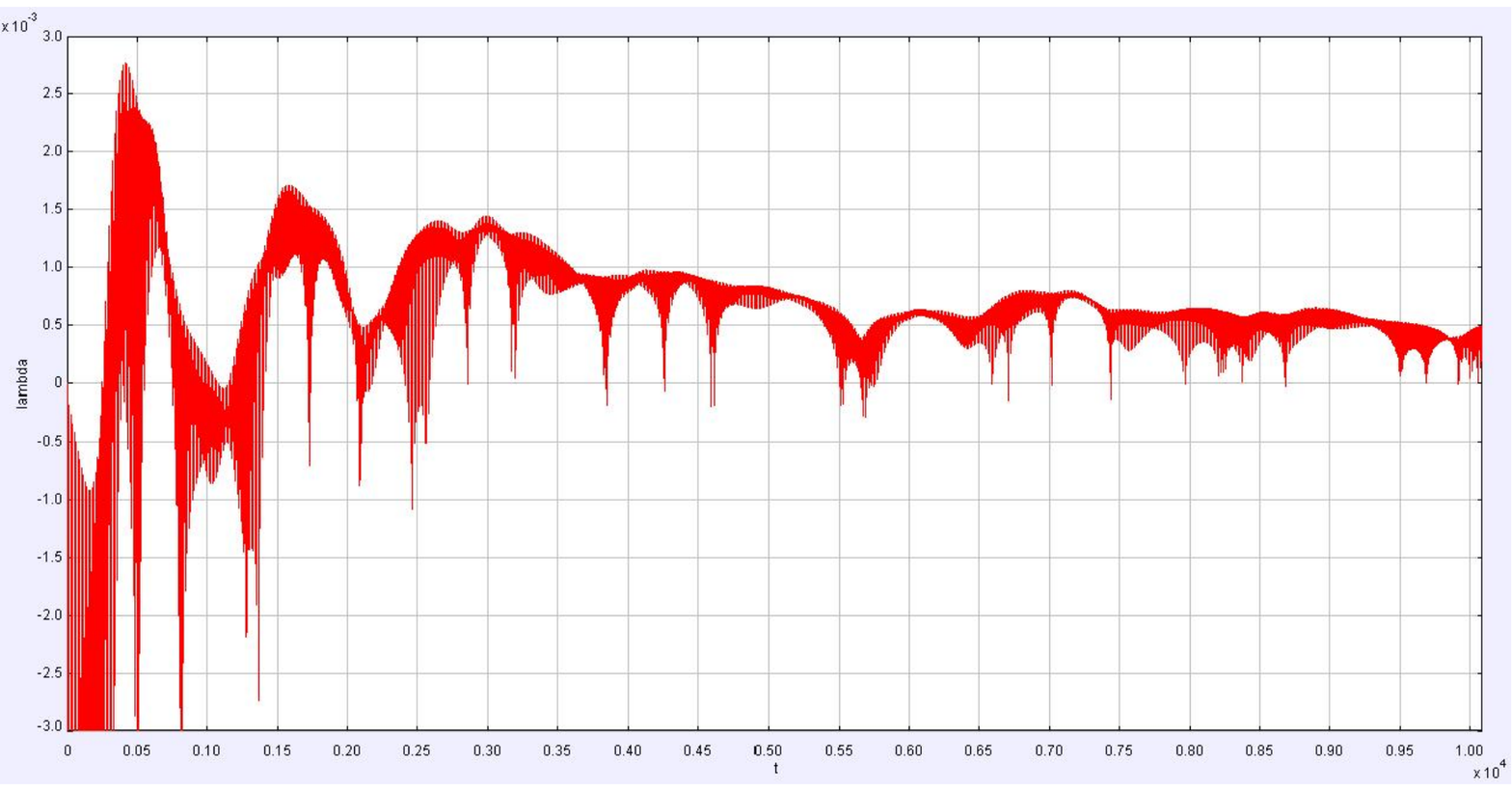}
\caption{Poincar\'e sections (left) and Lyapunov exponents (right) for the semiclassical equation of motion (\ref{qEOM}) with initial data $x_0=0.12$, $p_0=0.001$, $G_1=G_2=0.5$, $\Pi_1=\Pi_2=0$, and various values of  $\hbar^\prime$ as indicated.}
\label{Fig4}

\end{figure}

% -- Fig. 5
%----------------------------------
\begin{figure}[ht] \centering
\includegraphics*[width=7cm,height=5cm]{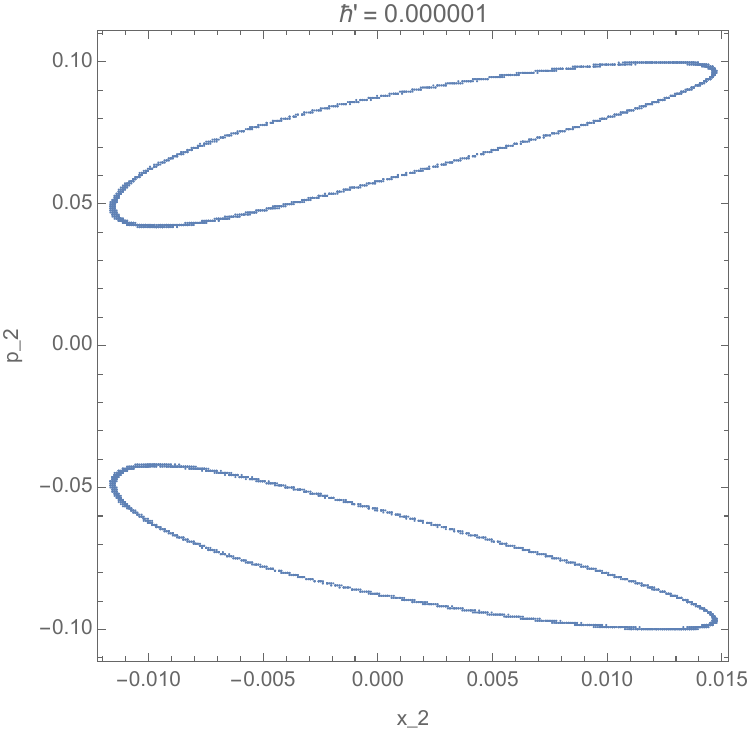}\includegraphics*[width=7cm,height=5.5cm]{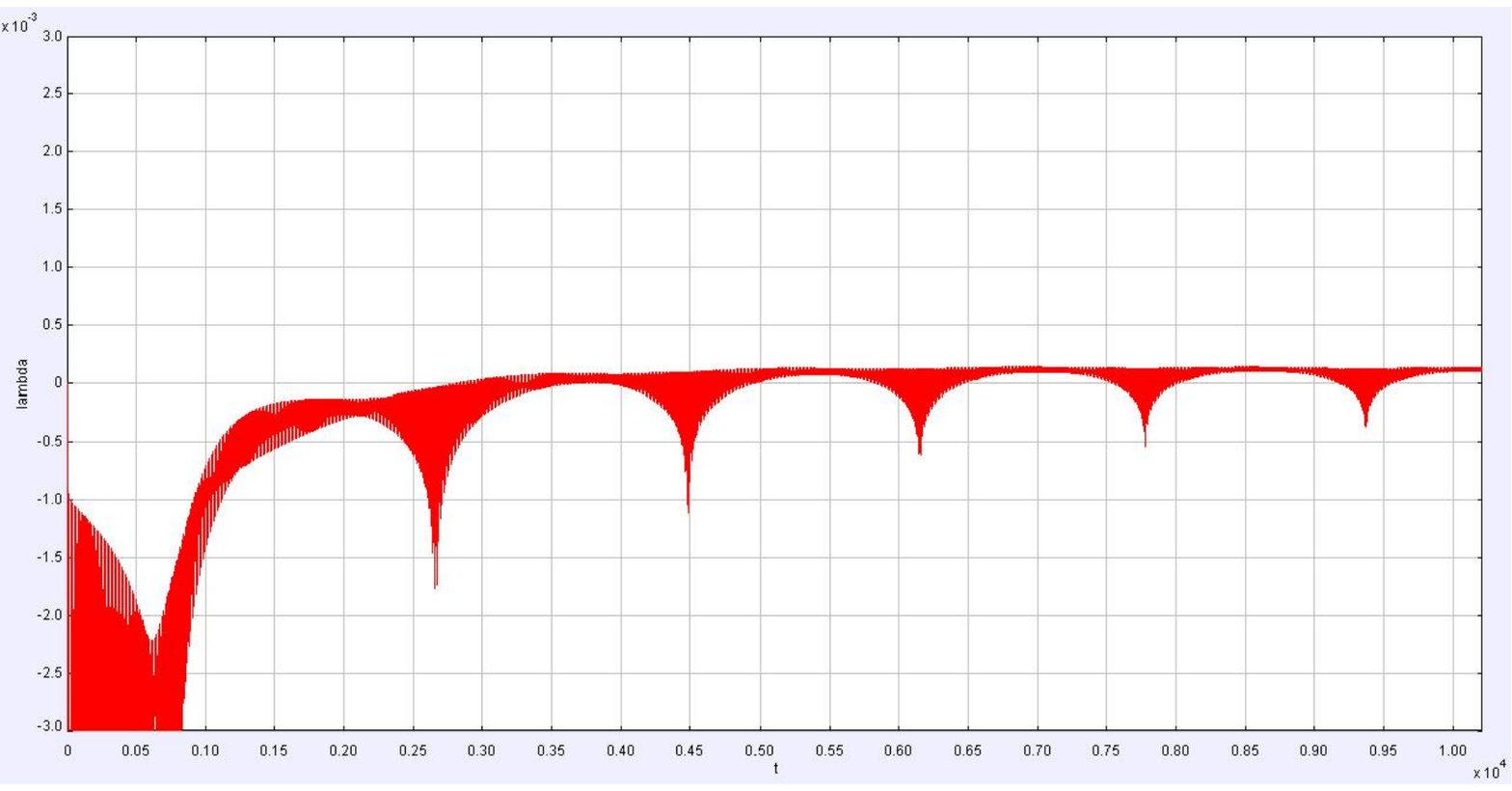}\\
\includegraphics*[width=7cm,height=5cm]{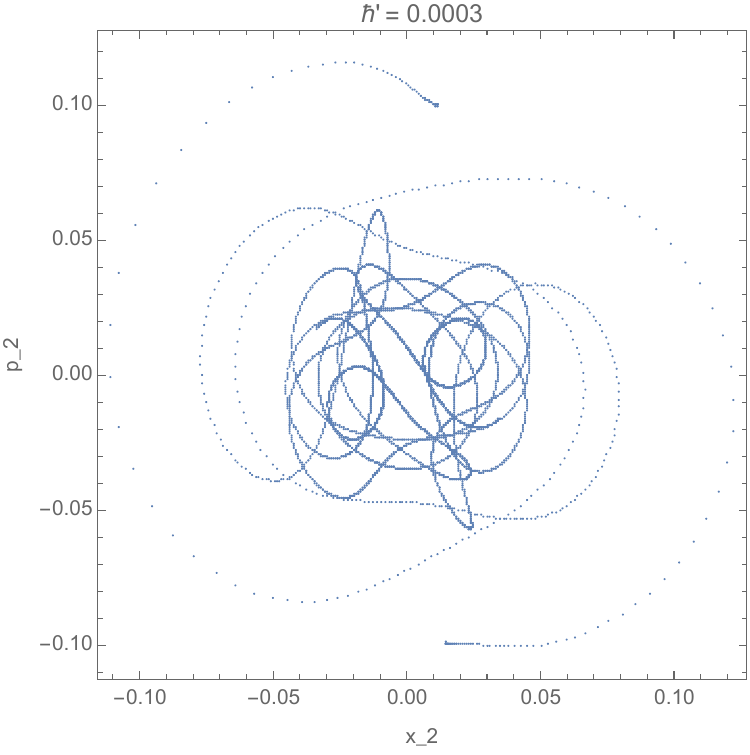}\includegraphics*[width=7cm,height=5.5cm]{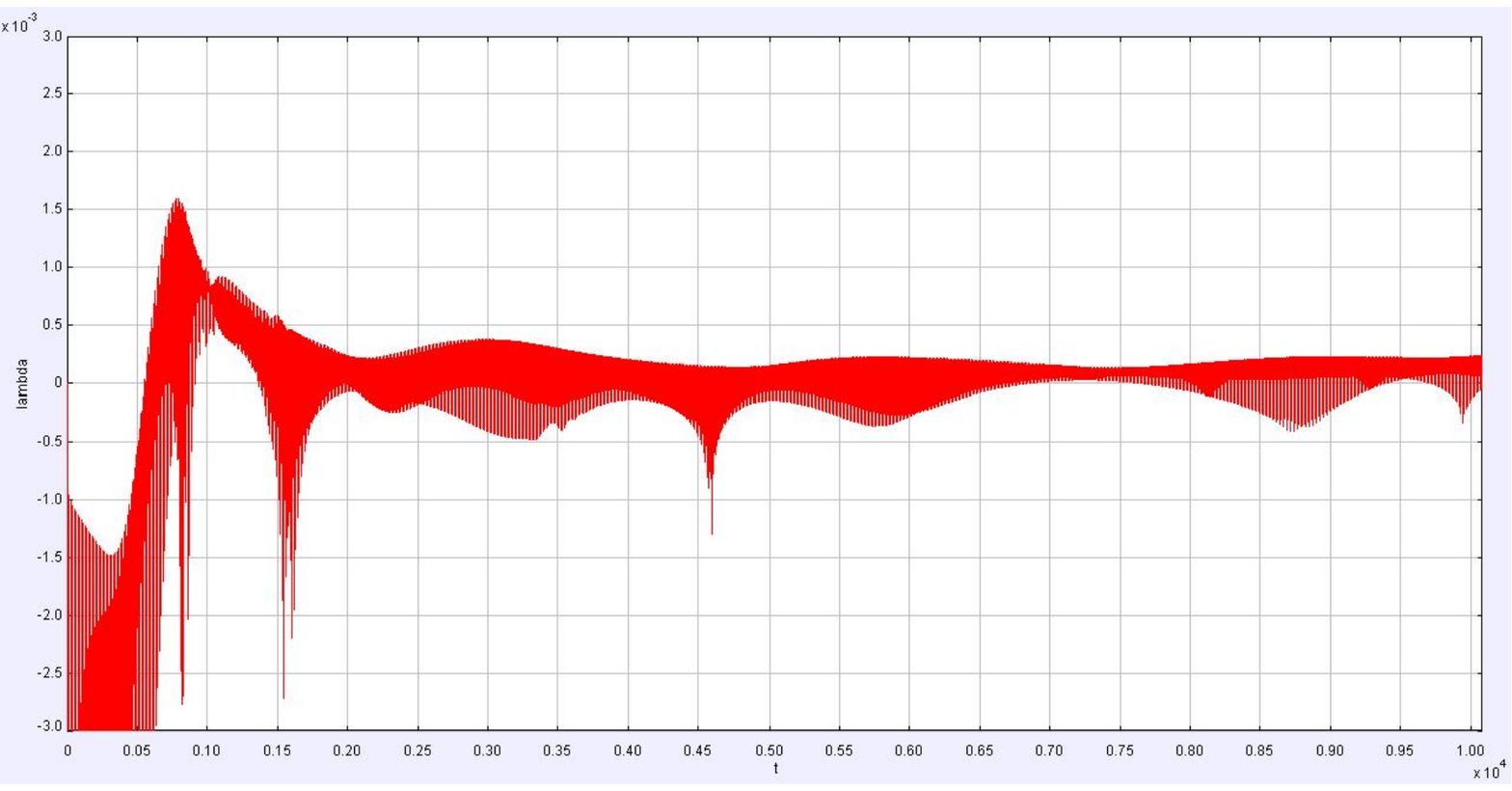}\\\includegraphics*[width=7cm,height=5cm]{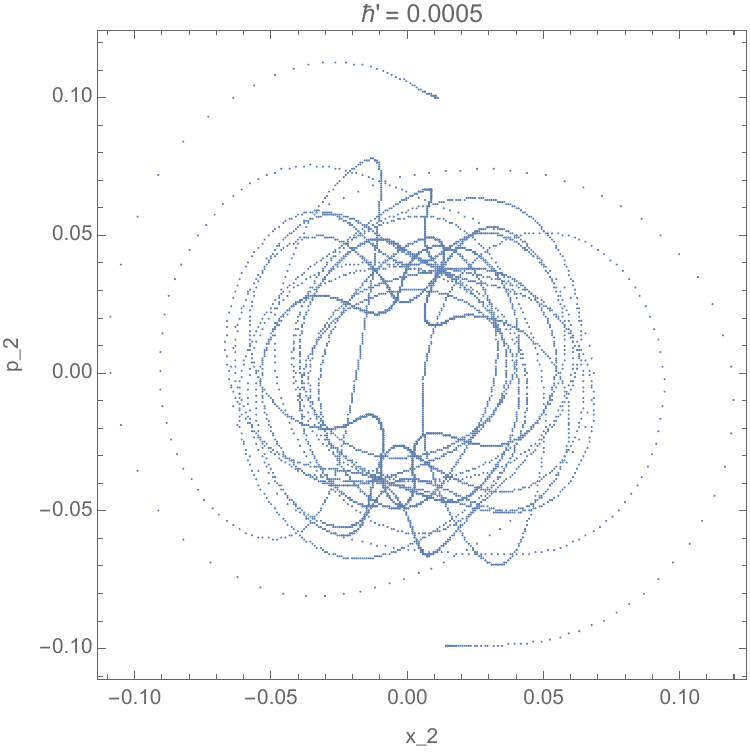}\includegraphics*[width=7cm,height=5.5cm]{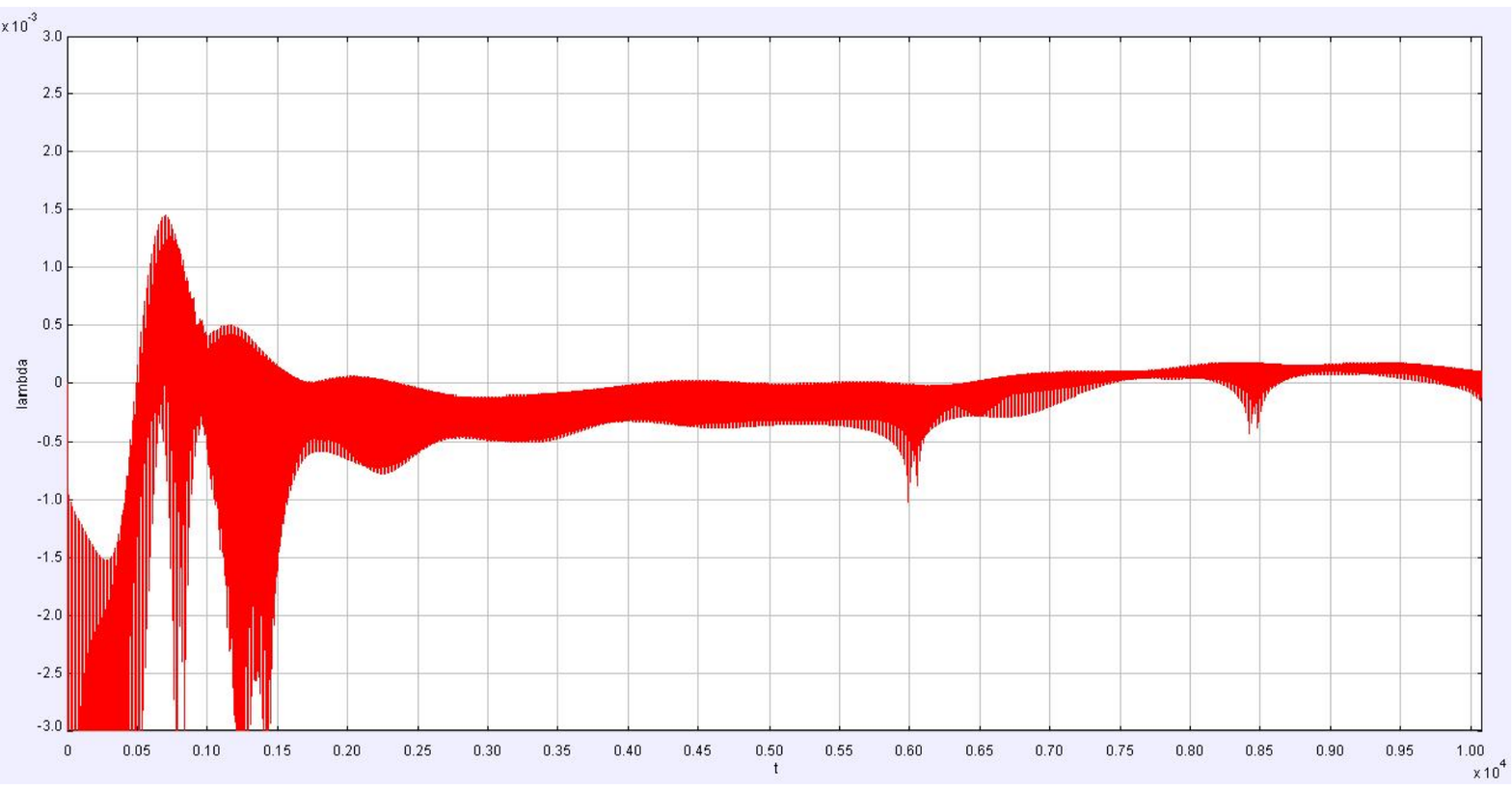}\\
\includegraphics*[width=7cm,height=5cm]{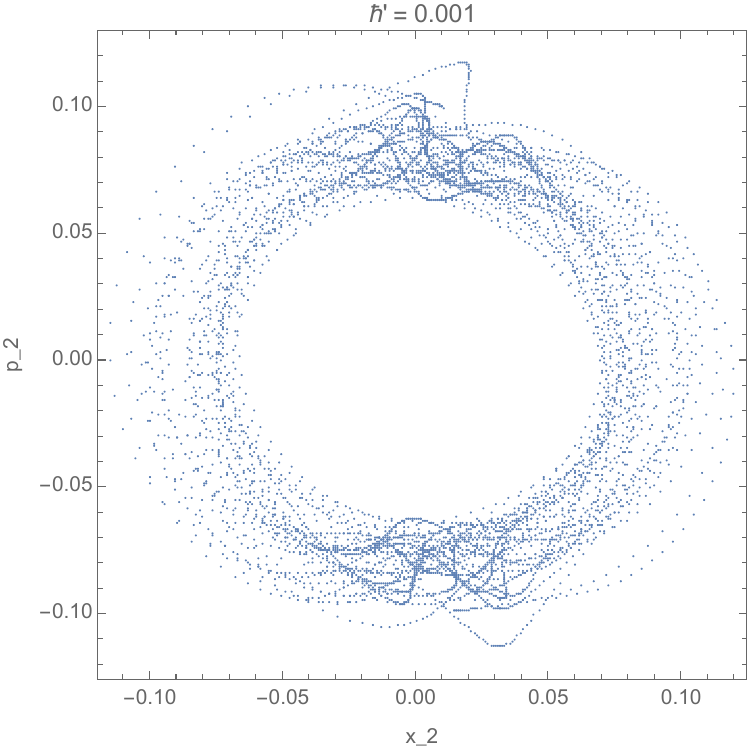}\includegraphics*[width=7cm,height=5.5cm]{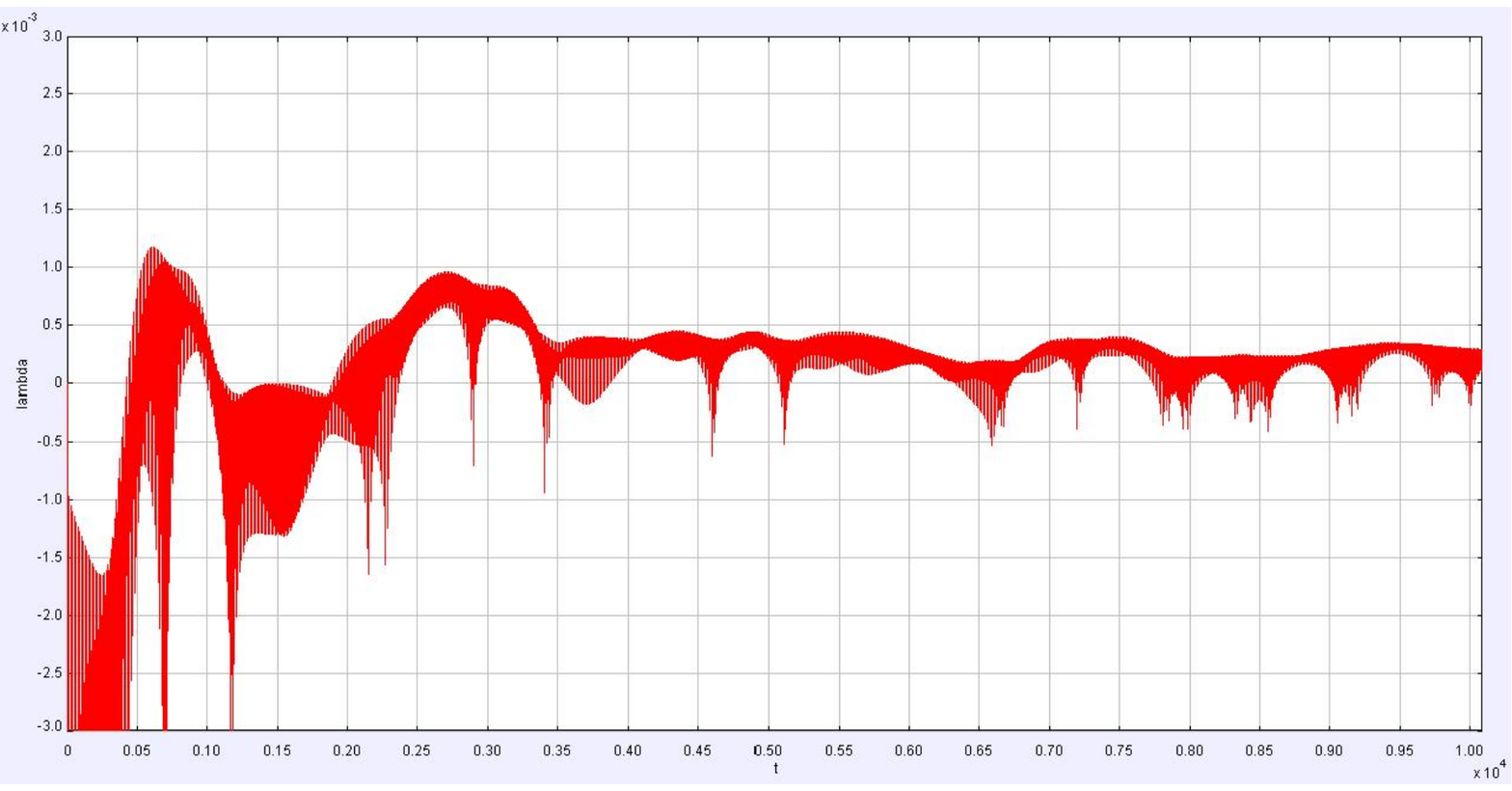}
\caption{Poincar\'e sections (left) and Lyapunov exponents (right) for the semiclassical equation of motion (\ref{qEOM}) with initial data $x_0=0.10$, $p_0=0.01$, $G_1=G_2=0.5$, $\Pi_1=\Pi_2=0$, and various values of  $\hbar^\prime$ as indicated.}
\label{Fig5}

\end{figure}

% -- Fig. 6
%----------------------------------
\begin{figure}[ht] \centering
\includegraphics*[width=7cm,height=5cm]{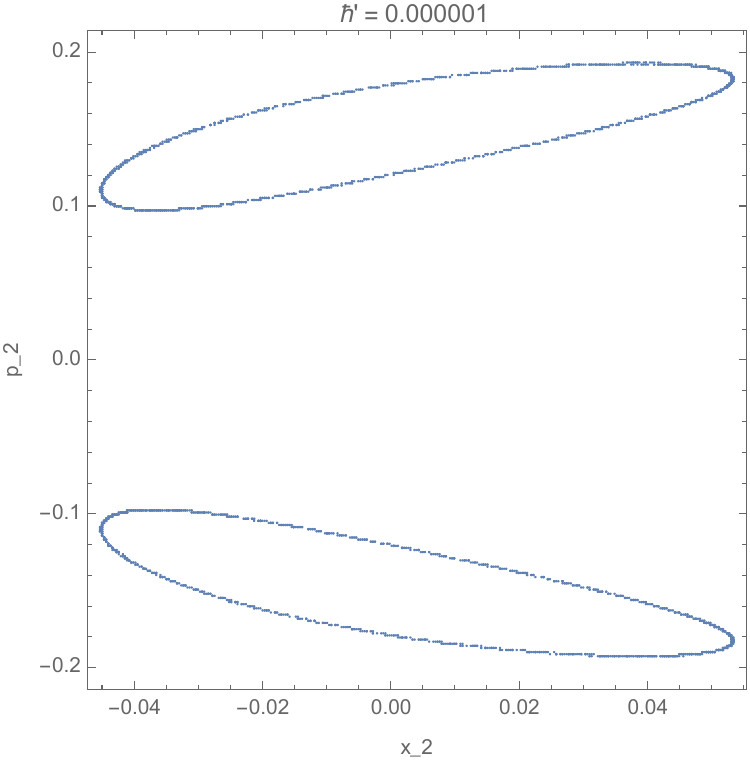}\includegraphics*[width=7cm,height=5.5cm]{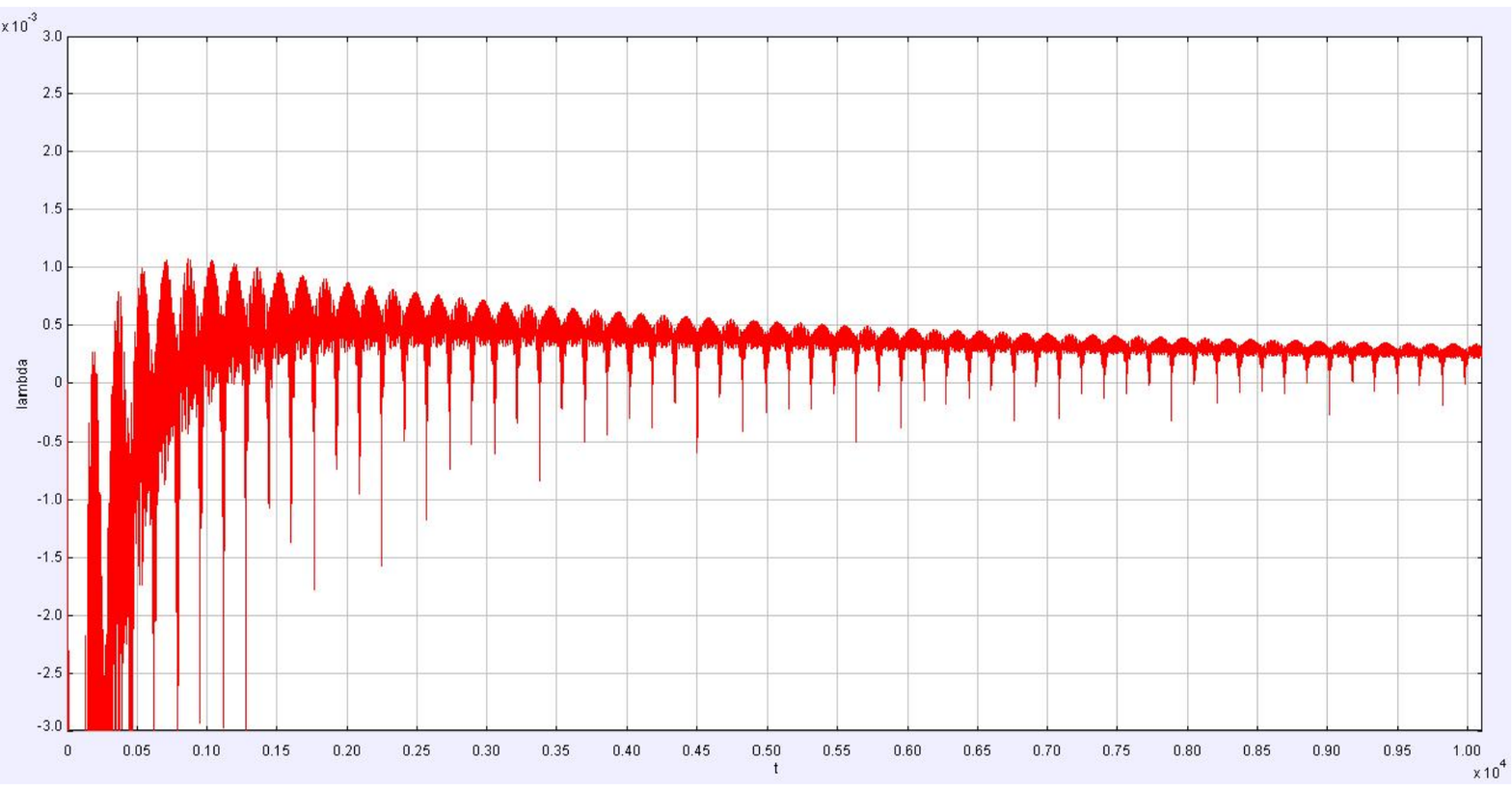}\\
\includegraphics*[width=7cm,height=5cm]{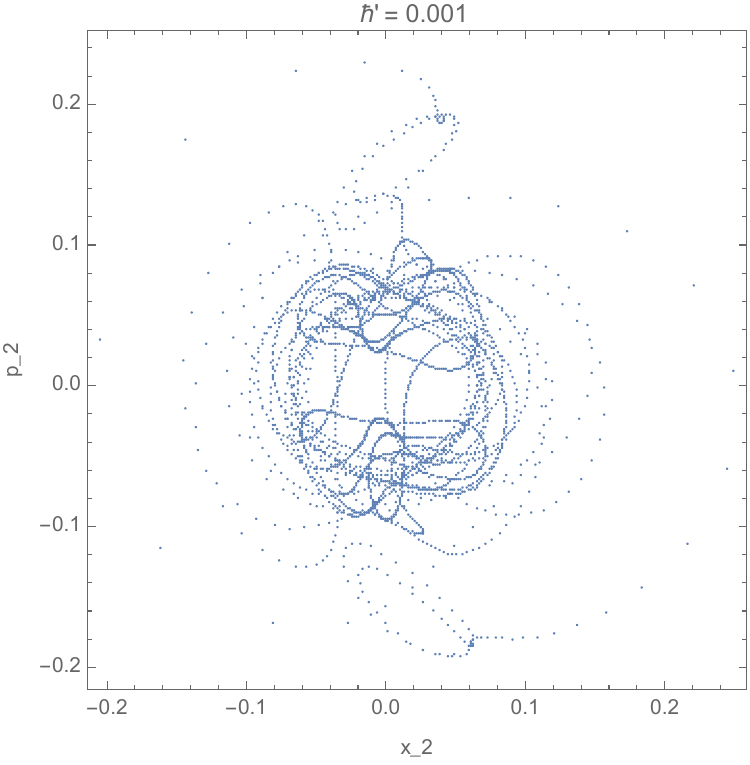}\includegraphics*[width=7cm,height=5.5cm]{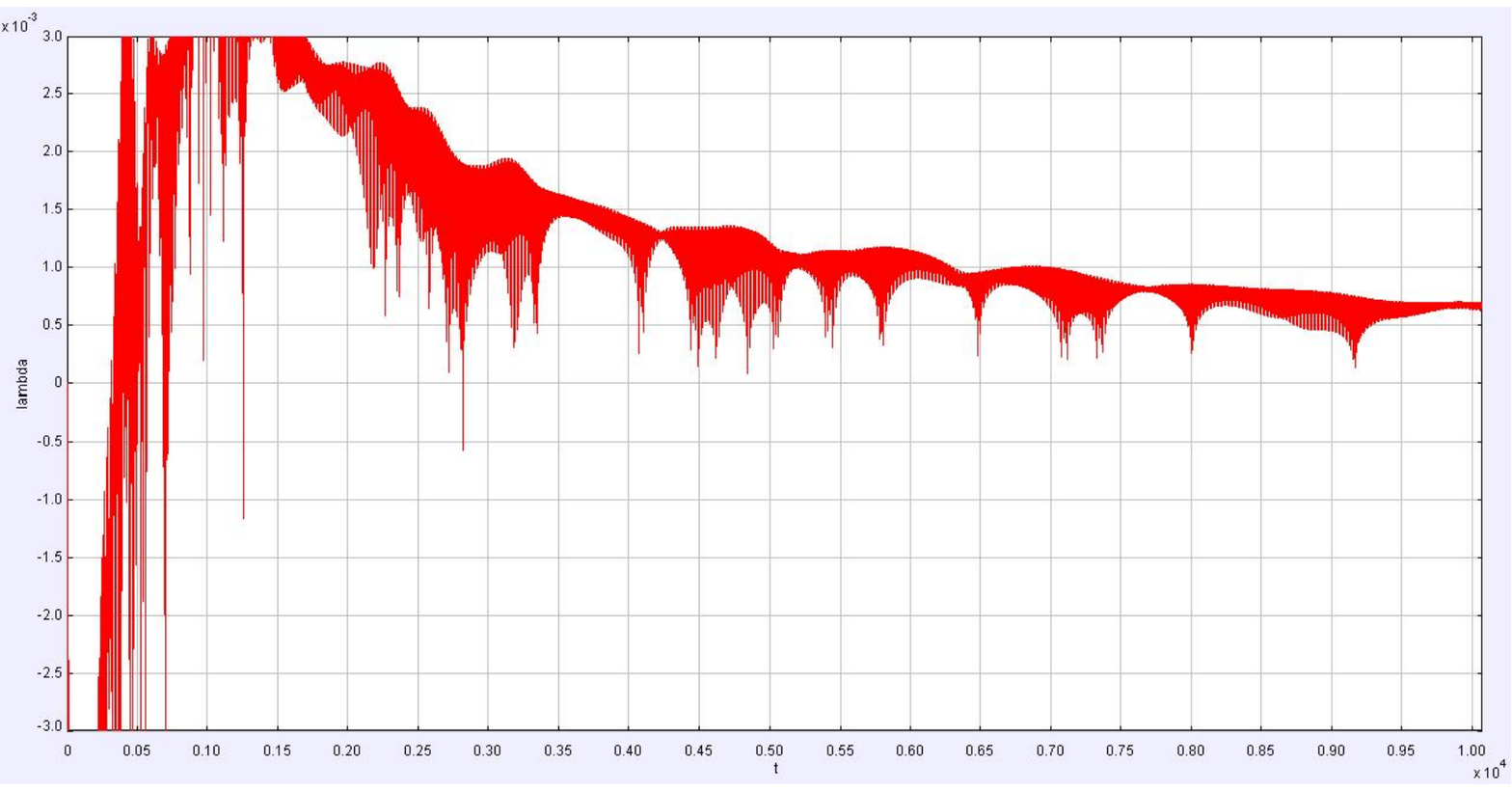}\\\includegraphics*[width=7cm,height=5cm]{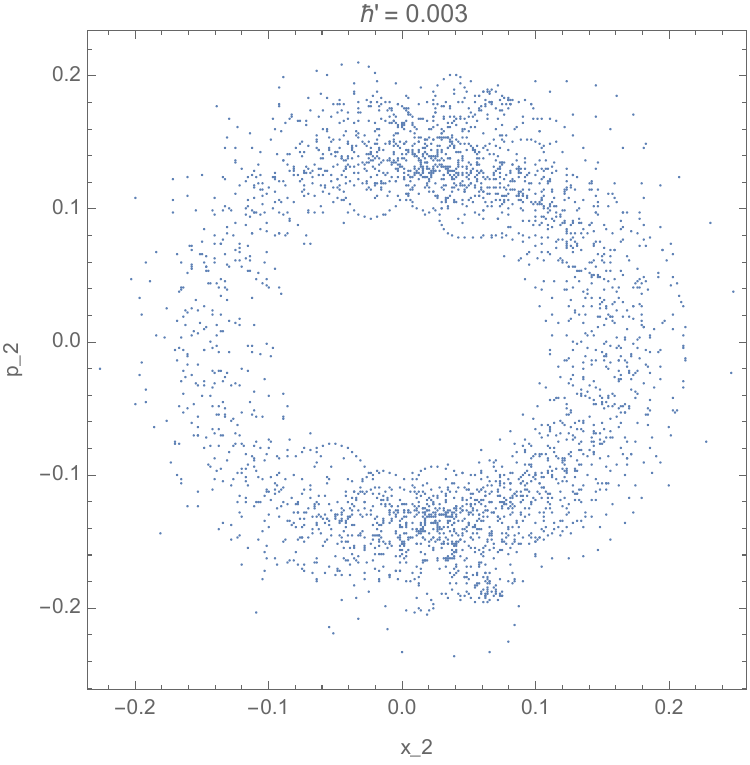}\includegraphics*[width=7cm,height=5.5cm]{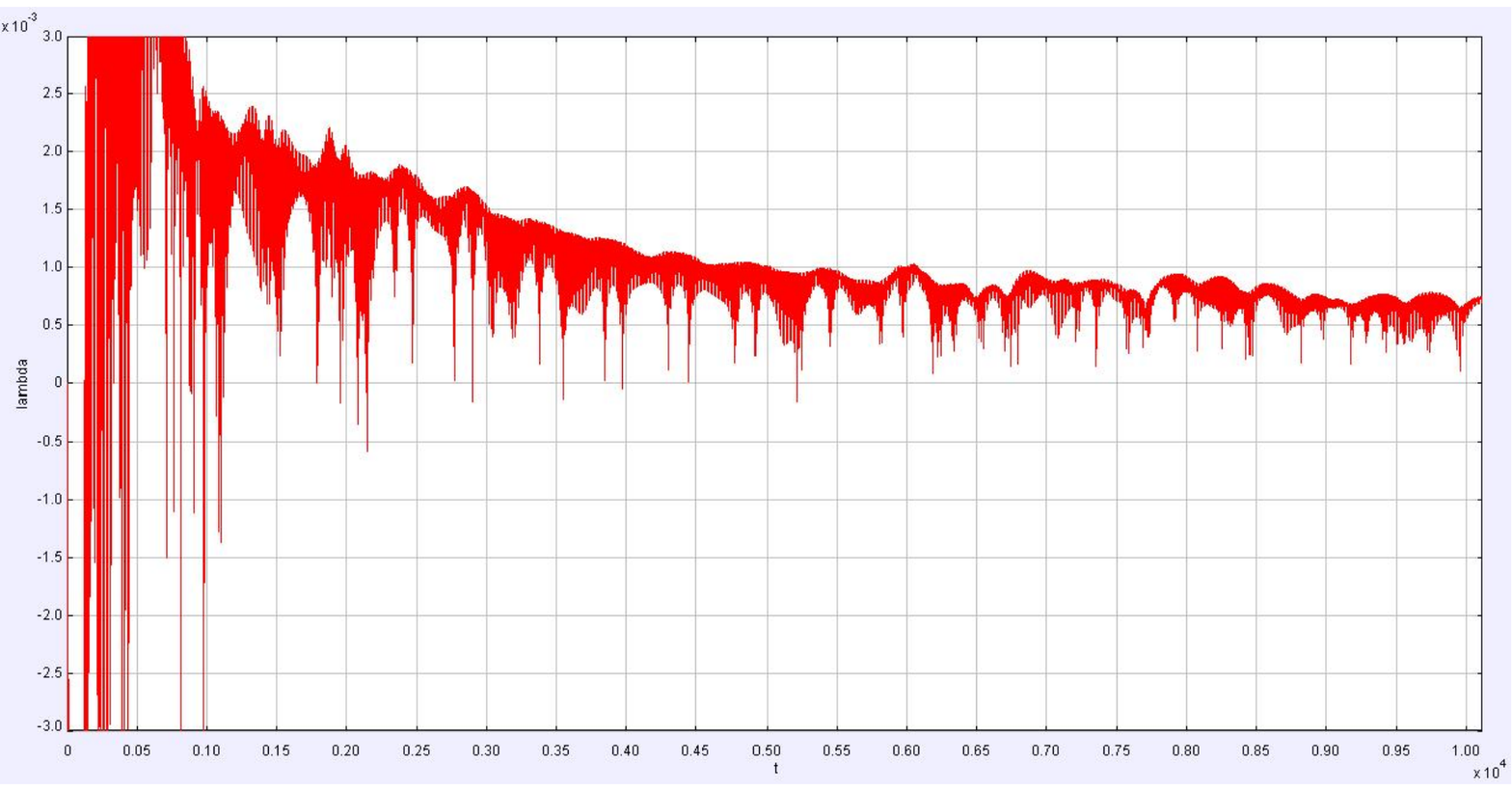}\\
\includegraphics*[width=7cm,height=5cm]{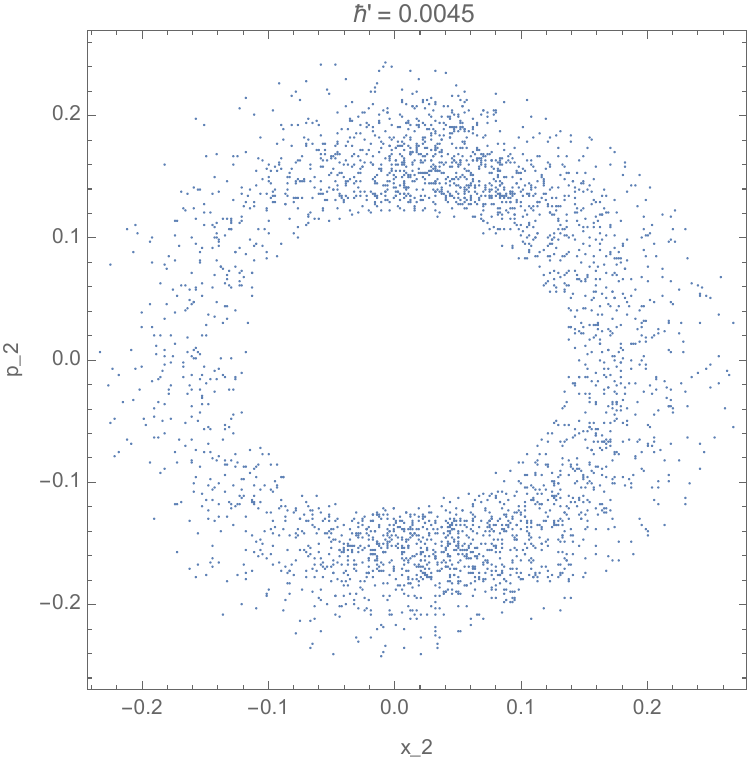}\includegraphics*[width=7cm,height=5.5cm]{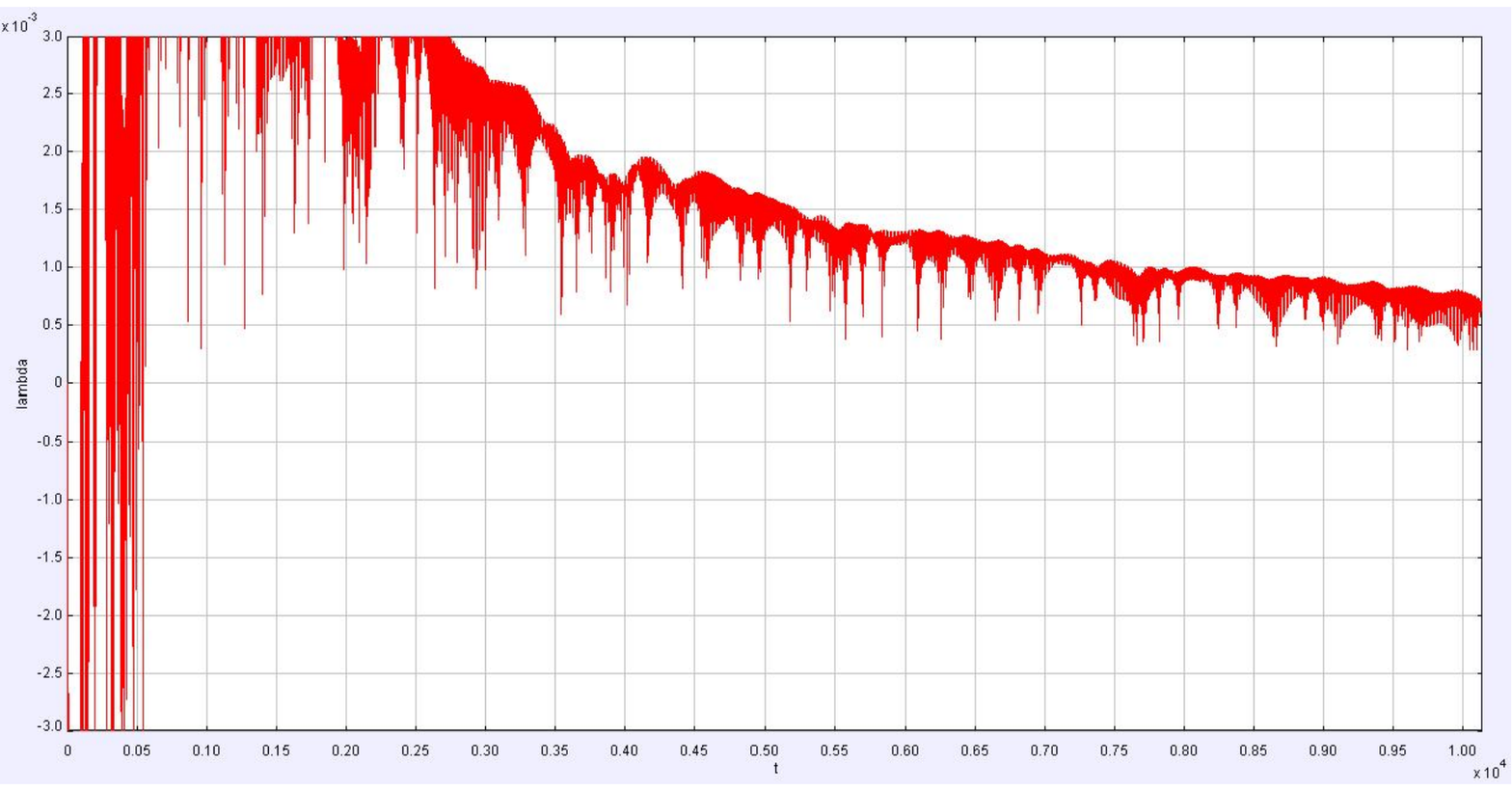}
\caption{Poincar\'e sections (left) and Lyapunov exponents (right) for the semiclassical equation of motion (\ref{qEOM}) with initial data $x_0=0.20$, $p_0=0.01$, $G_1=G_2=0.5$, $\Pi_1=\Pi_2=0$, and various values of  $\hbar^\prime$ as indicated.}
\label{Fig6}

\end{figure}

% -- Fig. 7
%----------------------------------
\begin{figure}[ht] \centering
\includegraphics*[width=7cm,height=5cm]{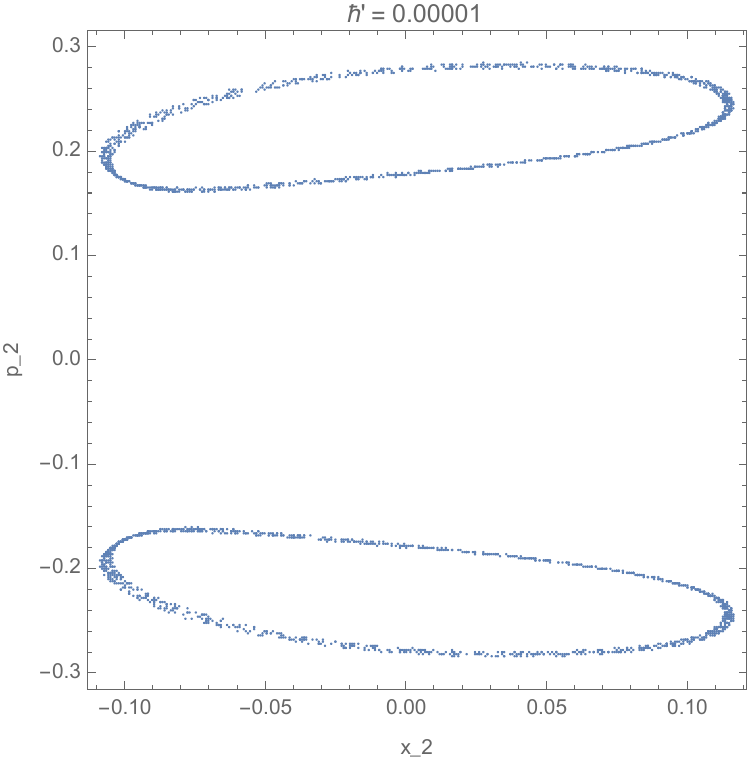}\includegraphics*[width=7cm,height=5.5cm]{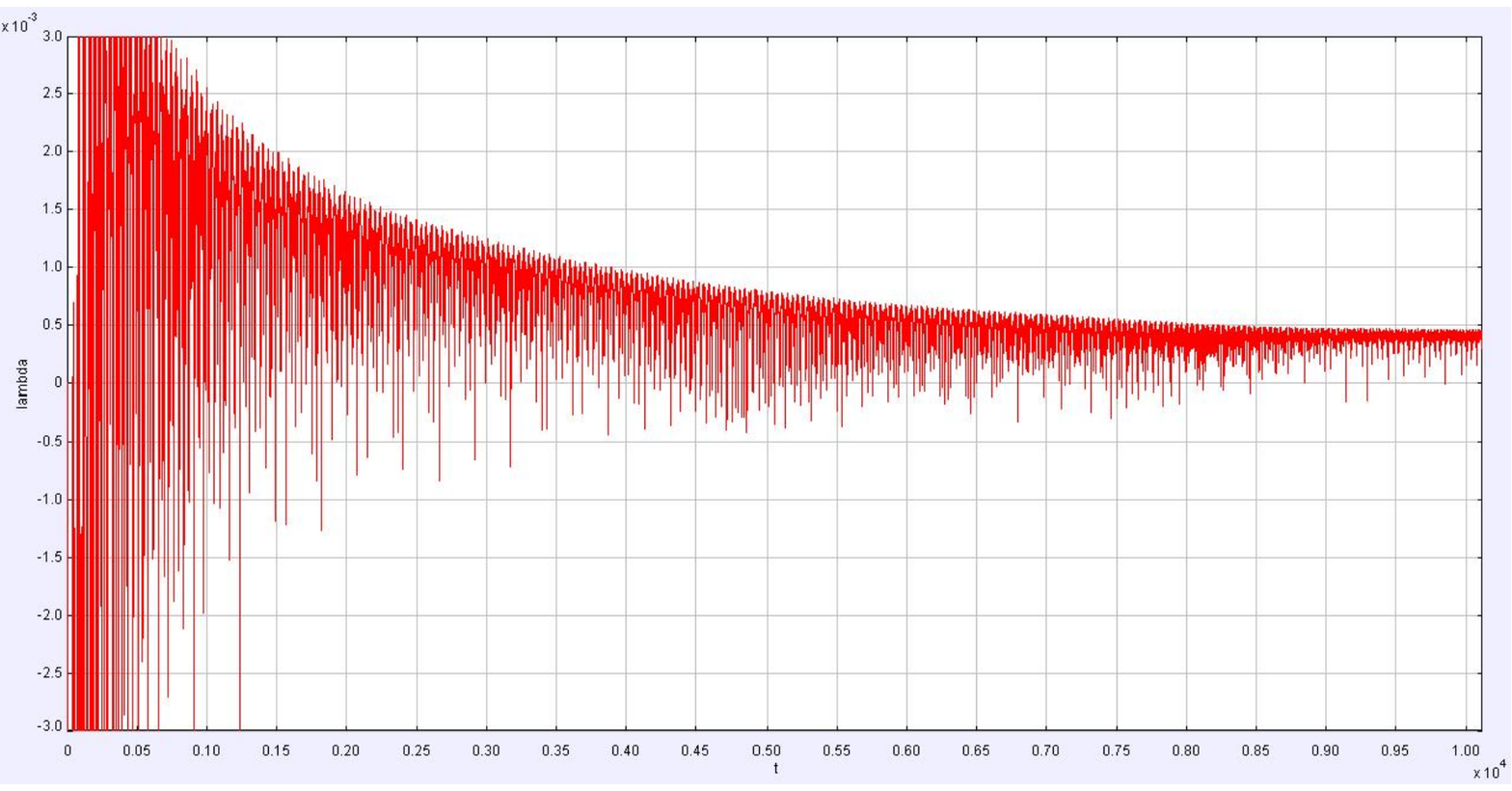}\\
\includegraphics*[width=7cm,height=5cm]{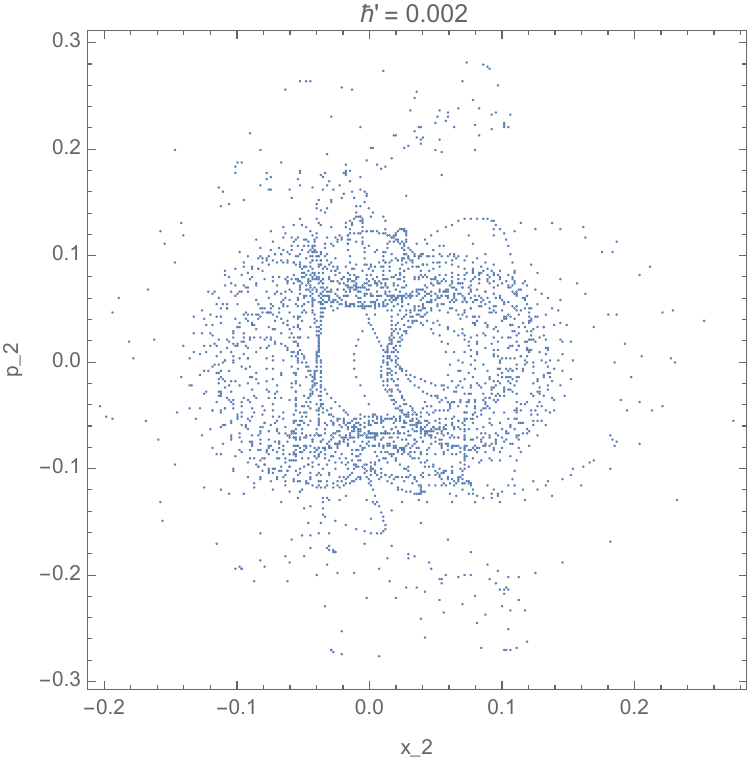}\includegraphics*[width=7cm,height=5.5cm]{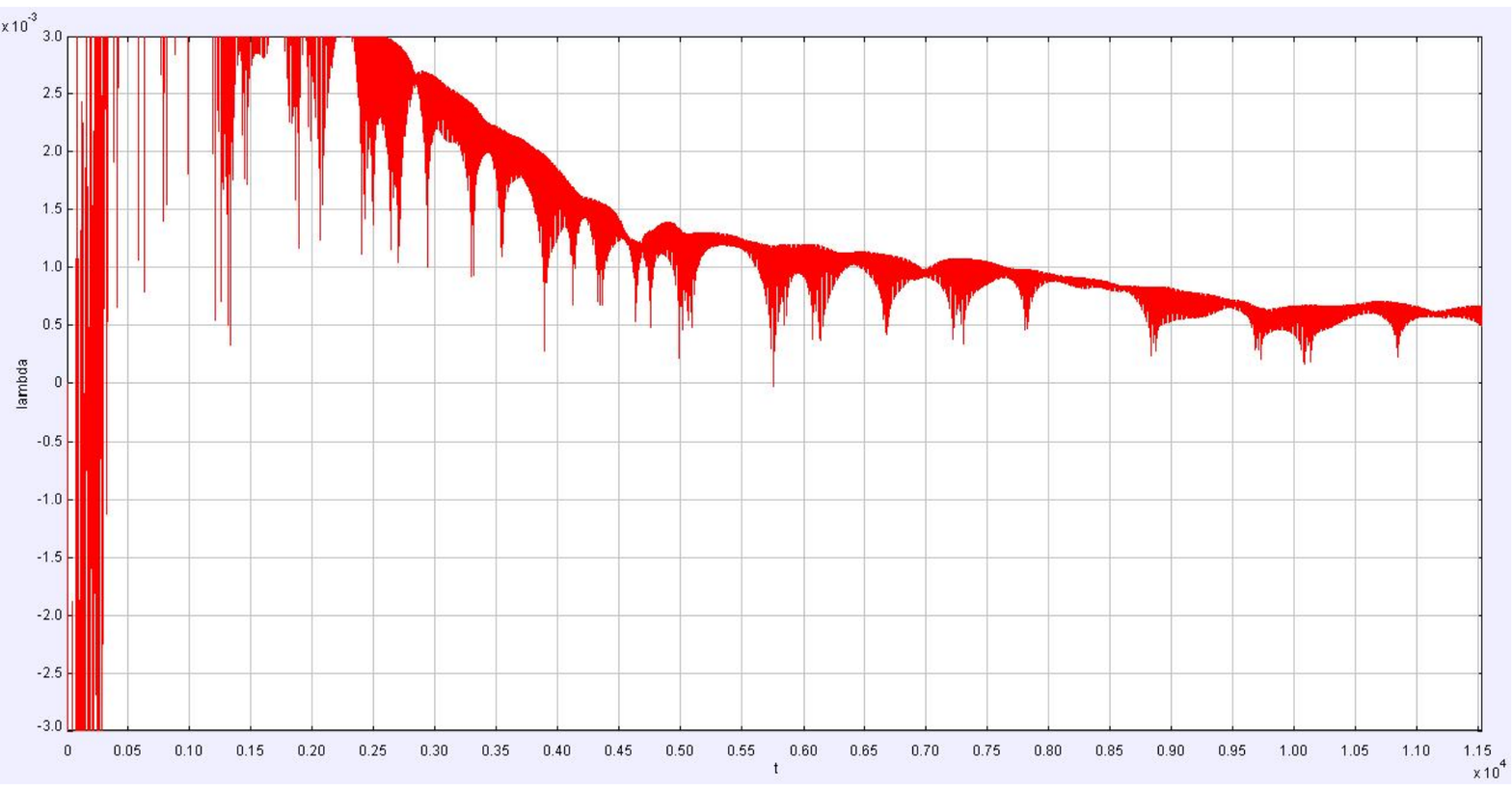}\\\includegraphics*[width=7cm,height=5cm]{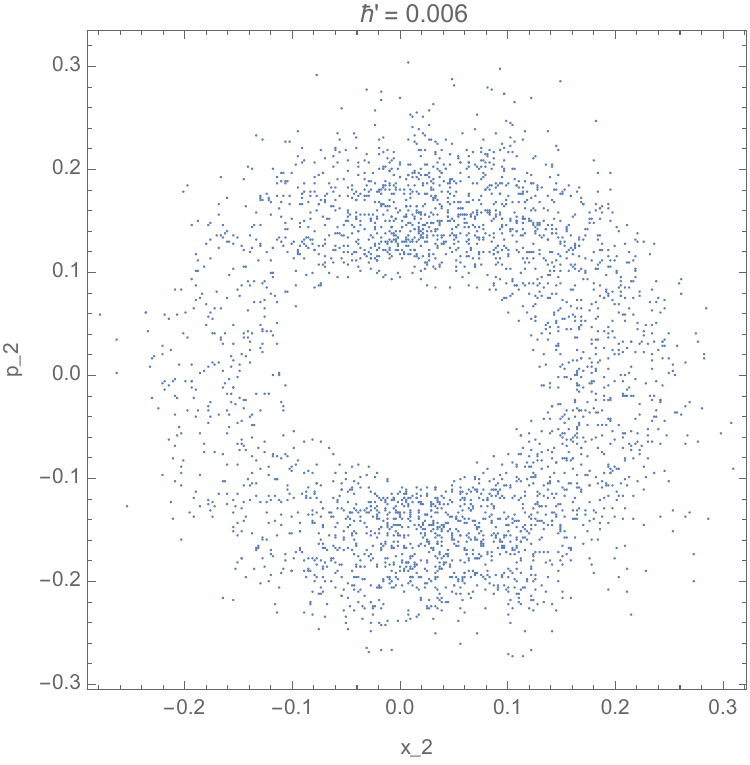}\includegraphics*[width=7cm,height=5.5cm]{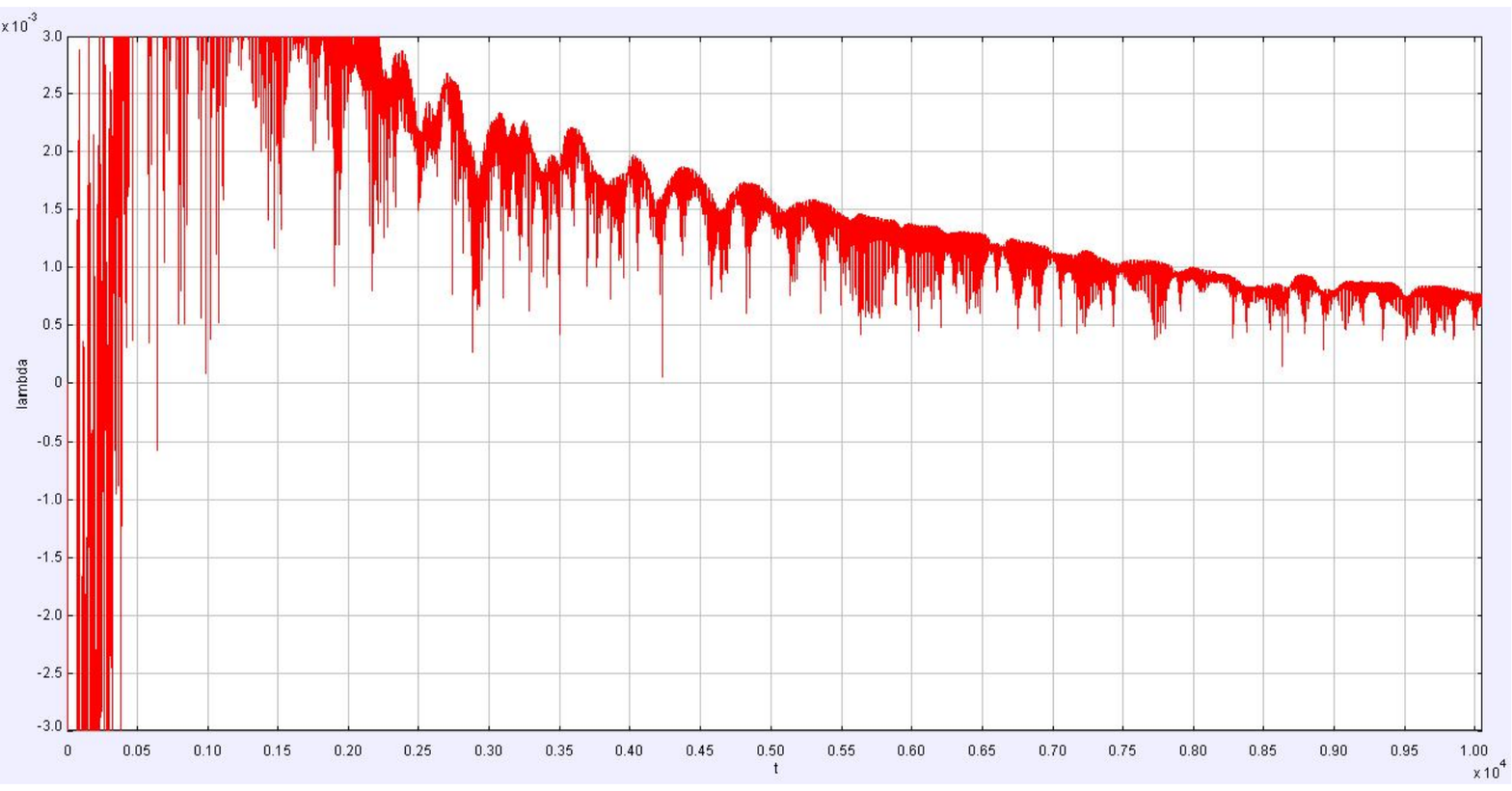}\\
\includegraphics*[width=7cm,height=5cm]{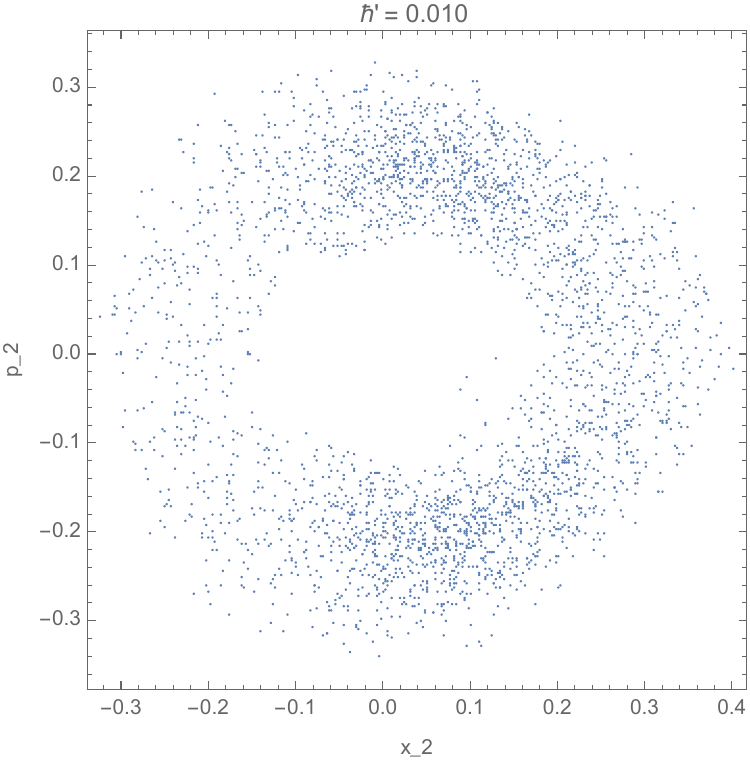}\includegraphics*[width=7cm,height=5.5cm]{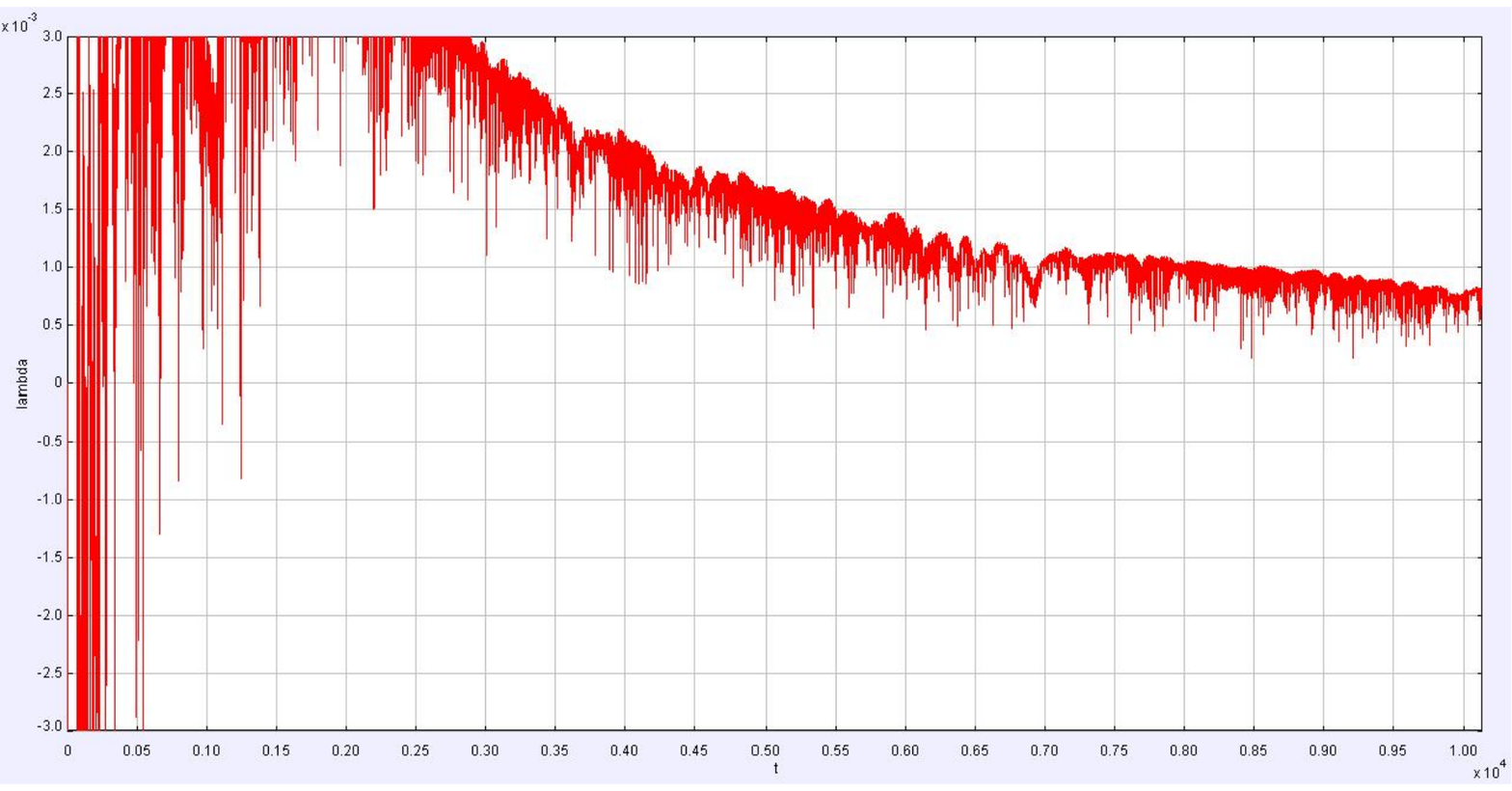}
\caption{Poincar\'e sections (left) and Lyapunov exponents (right) for the semiclassical equation of motion (\ref{qEOM}) with initial data $x_0=0.30$, $p_0=0.01$, $G_1=G_2=0.5$, $\Pi_1=\Pi_2=0$, and various values of  $\hbar^\prime$ as indicated.}
\label{Fig7}

\end{figure}

%--------------------------

\begin{thebibliography}{99}

%. Q many-body 

\bibitem{Bo}
F. Borgonovi, F. M. Izrailev, L. F. Santos, and V. G. Zelevinsky, 
Quantum chaos and thermalization in isolated systems of interacting particles, 
Phys. Rep. 626 (2016) 1.

\bibitem{Al}
L. D'Alessio, Y. Kafri, A. Polkovnikov, and M. Rigol, 
From quantum chaos and eigenstate thermalization to statistical mechanics and thermodynamics, 
Adv. Phys. 65 (2016) 239.

\bibitem{Su}
J. \u{S}untajs, J. Bon\u{c}a, T. Prosen, and L. Vidmar, 
Quantum chaos challenges many-body localization, 
arXiv:1905.06345 [cond-mat.str-el] (2019).

\bibitem{Ca}
Z. Cao, Z. Xu, and A. del Campo,
Probing quantum chaos in multipartite systems,
arXiv:2111.12475 [quant-ph] (2022).

%-- QFT ----

\bibitem{SS1}
S. H. Shenker and D. Stanford, 
Black holes and the butterfly effect,
J. High Energy Phys. 2014 (2014) 67.

\bibitem{SS2}
 S.H. Shenker and D. Stanford, 
 Stringy effects in scrambling, 
 J. High Energy Phys. 2015 (2015) 132.

\bibitem{MSS}
J. Maldacena, S. H. Shenker and D. Stanford, 
A bound on chaos,  
J. High Energy Phys.  2016 (2016) 106.
% [arXiv:1503.01409[hep-th]]

\bibitem{Co}
J.S. Cotler, G. Gur-Ari, M. Hanada, J. Polchinski, P. Saad, S.H. Shenker, D. Stanford, A. Streicher, and M. Tezuka,
Black holes and random matrices,
J. High Energy Phys. 2017 (2017) 118.

\bibitem{Ku}
S. Kundu, Extremal chaos, arXiv:2109.08693 [hep-th].

%--  H-atom chaos
\bibitem{FW} 
H. Friedrich and D. Wintgen, 
The hydrogen atom in a uniform magnetic field: An example of chaos, 
Phys. Rep. 183 (1989) 37.

%-- Billiard 
\bibitem{MK} 
S.W. McDonald and A.N. Kaufman, 
Spectrum and Eigenfunctions for a Hamiltonian with Stochastic Trajectories,
Phys. Rev. Lett. 42 (1979)1189.

%-- Haake
\bibitem{Ha} 
F. Haake, Quantum Signatures of Chaos, 3rd ed. (Springer, Berlin, 2010).

%-- Mahta
\bibitem{Me} 
M.L. Mehta, Random Matrices, 3rd ed. (Elsevier, San Diego, 2004).

%-- level dynamics
\bibitem{Dy} 
F.J. Dyson, Statistical theory of the energy levels of complex systems, 
J. Math. Phys. 3 (1962) 140-175.

%-- trace 
\bibitem{Gu} 
M.C. Gutzwiller, Chaos in Classical and Quantum Mechanics (Springer, New York, 1990).

%- Heller
\bibitem{He1}
 E.J. Heller, Time-dependent approach to semiclassical dynamics,
J. Chem. Phys. 62 (1975) 1544.

\bibitem{He2} 
E. J. Heller, Wavepacket dynamics and quantum chaology, in ``Chaos
and Quantum Physics'', 
Proceedings of the Les Houches Summer School, 1989 (North-Holland, Amsterdam, 1991).
 
%-- Q induces chaos
\bibitem{PS1} 
A.K. Pattanayak and W.C. Schieve,
Semiquantal Dynamics of Fluctuations: Ostensible Quantum Chaos,
Phys. Rev. Lett. 72 (1994) 2855.

\bibitem{Ber}
A. Bera, S. Dalui, S. Ghosh, and E.C. Vagenas, 
Quantum corrections enhance chaos: study of particle motion near a generalized Schwarzschild black hole,
arXiv:2109.00330 [gr-qc].

% --  Henon-Heiles
\bibitem{HH}  
M. H\'enon and C. Heiles, 
The applicability of the third integral of motion: Some numerical experiments, 
Astron. J. 69 (1964) 73.

\bibitem{LL} 
A. J. Lichtenberg and M. A. Lieberman, 
Regular and Stochastic Motion, 2nd ed.  (Springer-Verlag, New York, 1992).

\bibitem{Be}
M.V. Berry, Regular and irregular motion, in ``Topics in Nonlinear Dynamics" (ed. S. Jorna), 
Am. Inst. Phys.Conf. Proc. 46 (1978) 16.

\bibitem{Em}
C. Emanuelsson, Chaos in the H\'enon-Heiles system, Notes on Analytical Mechanics (FYGC04), University of Karlstad.

% -- QHH 
\bibitem{Noid} 
D.W.  Noid, M.L. Koszykowski, M. . Tabor, and R. A. Marcus,
Properties of vibrational energy levels in the quasi periodic and stochastic regimes,
J. Chem. Phys. 72 (1980) 6169.

%. -- Q destroy chaos in HH model -----
%  Effective potential - HH model
\bibitem{CS} 
L. Carlson and W. C. Schieve,
Chaos and quantum fluctuations in the H\'enon-Heiles and four-leg potentials,
Phys. Rev. A 40 (1989) 5896.

\bibitem{PS2} 
A.K. Pattanayak and W.C. Schieve,
EfFective potentials and chaos in quantum systems,
Phys. Rev. A 46 (1992) 1821.

%-- Dirac t-dep 
\bibitem{Di}
 P.A.M. Dirac, Appendix to Russian edition of: The principles of quantum mechanics, as cited by J. Frenkel, Ref.[28]; Proc. Camb.  Phil. Soc.  26 (1930) 376.

\bibitem{Fr} 
J. Frenkel, Wave mechanics: advanced general theory (Clarendon, Oxford, 1934) pp. 253, 435.

\bibitem{JK} 
R. Jackiw and A. Kerman, 
Time-dependent variational principle and the effective action,
Phys. Lett. {\bf A 71}, 158 (1979).

\bibitem{TF} Y. Tsue and Y. Fujiwara, 
Time-Dependent Variational Approach in Terms of Squeezed Coherent States,
Prog. Theor. Phys. 86 (1991) 443.

\bibitem{Ho}  C.-L. Ho and C.-I. Chou, Simple variational approach to
quantum Frenkel-Kontorova model, Phys. Rev. E 63 (2001) 016203.

\bibitem{BE}
{
T. Blum, H.-Th. Elze, Semiquantum chaos in the double-well,
 Phys. Rev. E 53 (1996) 3123.
 }
 
 \bibitem{E1}{
 H.-Th. Elze, 
Quantum decoherence, entropy and thermalization in strong interactions at high energy,
Nucl. Phys. B436 (1995) 213.
}

\bibitem{E2}
H.-Th. Elze,
Entropy, quantum decoherence and pointer states in scalar parton fields,
Phys. Lett. B369 (1996) 295.


\end{thebibliography}
\end{document}